\documentclass[twocolumn,showpacs,preprintnumbers,amsmath,amssymb]{revtex4-1}
\usepackage{graphicx}
\usepackage{dcolumn}
\usepackage{bm,color}

\newcommand{\be}{\begin{eqnarray}}
\newcommand{\ee}{\end{eqnarray}}

\begin{document}

\preprint{YGHP-15-05}

\title{Short range inter-vortex interaction and interacting dynamics of half-quantized vortices in two-component Bose-Einstein condensates}

\author{Kenichi Kasamatsu$^{1}$}
\author{Minoru Eto$^{2}$}
\author{Muneto Nitta$^{3}$}
\affiliation{
$^1$Department of Physics, Kinki University, Higashi-Osaka, 577-8502, Japan \\
$^2$Department of Physics, Yamagata University, Yamagata 990-8560, Japan \\
$^3$Department of Physics, and Research and Education Center for Natural 
Sciences, Keio University, Hiyoshi 4-1-1, Yokohama, Kanagawa 223-8521, Japan }

\date{\today}

\begin{abstract}
We study the interaction and dynamics of two half-quantized vortices 
in two-component Bose-Einstein condensates. 
Using the Pad\'{e} approximation for the vortex core profile, we calculate 
the intervortex potential, whose asymptotic form for a large distance has been 
derived by Eto {\it et al.} [Phys. Rev. A, \textbf{83}, 063603 (2011)]. 
Through numerical simulations of the two-dimensional Gross-Pitaevskii equations, 
we reveal different kinds of dynamical trajectories of the vortices depending on 
the combinations of signs of circulations and the intercomponent density coupling. 
Under the adiabatic limit, we derive the equations of motion 
for the vortex coordinates, in which the motion is caused by the balance between Magnus force 
and the intervortex forces. The initial velocity of the vortex motion 
can be explained quantitatively by this point vortex approximation, but understanding the long-time 
behavior of the dynamics needs more consideration beyond our model. 

\end{abstract}

\maketitle

\section{Introduction} \label{intro}
Exotic topological defects are interesting subjects in various physical area 
from condensed matter physics to high energy physics \cite{Volovikbook}. 
Among them these topological defects, 
exotic vortices, e.g., half-quantized vortices (HQVs) 
or more generally fractional vortices \cite{Babaev:2002}, 
can appear in the systems of multicomponent 
order parameters.
HQVs have been discussed in various area of physics 
\cite{Mineev}, and recently they have been observed 
experimentally in triplet superconductors \cite{Jang}, 
exiton-polariton condensates \cite{Lagoudakis,Manni}, 
ultracold atomic gas Bose-Einstein condensates (BECs)  \cite{Seo}, 
and helium 3 superfluids \cite{Autti:2015}.
Among them 
quantized vortices in multicomponent 
BEC are one of the important subjects \cite{Leonhardt,Kasarev}. 

Quantized vortices in cold atomic BECs  have been studied thoroughly 
since its experimental realization \cite{Fetterrev}. 
The recent progress in this subject can be seen in development of the 
techniques to nucleate the vortices and to detect the real time vortex dynamics. 
The group of Amherst College demonstrated the 
observation of real time evolution of a single vortex, a vortex dipole, and a cluster of a few vortices 
in a single-component BEC \cite{Freilich,Middelkamp,Navarro}. These experimental observation was 
compared with the simple intuitive theories of the point vortex approximation, 
where the vortex motion is described by fundamental equations 
of the vortex positions and the interaction between point vortices.  

Two-component BECs are a simplest example of the
multicomponent condensates and have also attracted much
interest to study the novel phenomena not found in a single component BEC. 
The mass circulation of two-component BECs is fractionally quantized 
when their mass densities are different. 
Such a quantized vortex in two-component BECs has a composite structure, 
where a vortex core in one component is filled by the density of the 
other component \cite{Kasarev,Jezek,Mueller,Kasaprl,Catelani}. This vortex structure was created experimentally 
through coherent interconversion between two components \cite{Matthews}. 

Recently, the asymptotic form of the interaction between two HQVs in two-component 
BECs was derived analytically, having the form $\sim (\ln R)/R^2$ for HQVs separated 
by a large distance $R$ \cite{Eto}. 
This $R$-dependence is different from that of the vortex-vortex interaction $\sim \ln R$ in a 
single-component BEC \cite{Pethickbook}. Since the two components interact only 
through the density-density coupling, 
the vortex in one component does not directly feel the circulation of the vortex 
in the other component. 
This fact results in an indirect interaction of the HQVs, where the filling component 
of each vortex core is affected by the circulation created by the vortex 
in the same component, dragging the vortex in which it is filled by the intercomponent 
density-density coupling. 
While the vortex interaction in a scalar BEC depends on the circulations of each vortex, 
the interaction for the HQVs does not depend on the sign of the circulation 
but on the sign and magnitude of the intercomponent coupling. 
Aftalion {\it et al}. studied the equilibrium properties of multi-vortex systems in two-component 
BECs using the above asymptotic form \cite{Aftalion}, and 
some studies have discussed the vortex-vortex interaction beyond the asymptotic regime 
\cite{Shirley,Dantas}.

A next question naturally arises as how two interacting HQVs behave temporally.
Although the real time dynamics of two vortices in two-component BECs 
was studied numerically by \"{O}hberg and Santos \cite{OS2002}, they considered 
only particular situations in a harmonically trapped system. 
Nakamura {\it et al.,} studied the $n$-vortex dynamics in multicomponent BECs through 
the variational analysis and the resulting equations of motions for the vortex positions 
are shown to include the terms of inertial vortex mass \cite{Nakamura}, which are usually 
neglected in the case of vortex motion in a scalar BEC. 

In this paper, we study the real time evolution of the two interacting HQVs in 
homogeneous two-component BEC. 
We find that depending on not only the vortex interaction but also the signs of circulation of the HQVs, 
the vortex motions exhibit interesting nontrivial trajectories. 
We consider the vortex dynamics for two cases 
(i) one of two vortices is placed in either of the two components and 
(ii) two vortices are placed in the same component. 
For the case (i), even though the intervortex interaction is independent of the sign of circulation, 
the dynamical trajectories of two vortices are different for the signs as well as for the 
intercomponent coupling strength. 
In the case (ii), the dynamics also has nontrivial dependence on the 
intercomponent coupling. 
The initial stage of the dynamics can be explained by the classical equations of motions for 
the point vortices, where it is important to take account of the short-range property 
of the vortex-vortex interaction. 

This paper is organized as follows. Section~\ref{basic} is devoted for representing 
the basic formulation of the following numerical simulations 
and the analytical calculations. In Sec.~\ref{case2}, we consider the interaction and dynamics 
of two vortices, in which one of these is put on either of the two components. 
We study the interacting dynamics by directly solving the two-dimensional Gross-Pitaevskii (GP) 
equations for two-component BECs. We interpret the numerical results by 
handling the point vortex approximation, finding that the proper treatment of the 
short range force between the two vortices can explain the short time behavior 
of the numerical results. 
We also consider the dynamics for two vortices which are put on the one of 
the two components in Sec.~\ref{case1} using similar treatments in Sec.~\ref{case2}. 
Summary and discussions are presented in Sec.~\ref{summary}. 

\section{Vortices in two-component Bose-Einstein condensates} \label{basic}
In this section, we give the theoretical formulation to study the vortex dynamics 
in two-component BECs. Usually, atomic condensates are trapped by harmonic 
traps, which give rise to the density inhomogeneity with an inverted parabolic 
form. Such an inhomogeneity has significant influence to vortex dynamics \cite{Fetterrev}. 
Since we are willing to consider the intrinsic dynamics caused by the vortex-vortex 
interaction, we restrict ourselves to the homogeneous system. 
This situation is realistic because a recent experiment demonstrated a realization 
of degenerate Bose gases in a uniform box potential \cite{Gaunt}. 

\subsection{The Gross-Pitaevskii model}
We start with the Lagrangian for homogeneous two-component BEC systems
\begin{align}
 L &= \int d \mathbf{r} \biggl\{ \sum_{i=1,2} \biggl[ -\frac{i\hbar}{2} ( \Psi_i \dot{\Psi}_i^{\ast} -  \Psi_i^{\ast} \dot{\Psi}_i  ) 
- \frac{\hbar^2}{2m_i} | \nabla \Psi_i|^2 \nonumber \\ 
&-\mu_i |\Psi_i|^2 - \frac{g_{i}}{2} |\Psi_i|^4 \biggr] - g_{12} |\Psi_1|^2  |\Psi_2|^2  \biggr\},
\label{eq:GP_lag}
\end{align}
where $\Psi_i$ is a condensate wave function 
of the $i$-th component ($i=1,2$) with the mass $m_i$ and the chemical potential $\mu_i$.
The coupling constants $g_1,g_2$ and $g_{12}$ represent 
the atom-atom interactions proportional to the $s$-wave scattering length; 
the $\Psi_1$ and $\Psi_2$ components repel or attract for $g_{12}>0$ or $g_{12} < 0$, respectively. 
We restrict ourselves to the miscible situation, where
the equilibrium densities without vortices are $n_{10} = (\mu_1 g_2 - \mu_2 g_{12})/(g_1 g_2 - g_{12}^2)$ 
and $n_{20} = (\mu_2 g_1 - \mu_1 g_{12})/(g_1 g_2 - g_{12}^2)$. 
To normalize the equation, we introduce the scales of length and time 
by $\xi = \sqrt{ \hbar^2/ 2 m_1 \mu_1}$ and $\hbar/\mu_1 = \tau$. 
Replacing $t/\tau \to t$, $\mathbf{r}/\xi \to \mathbf{r}$, and 
$\Psi_i = \sqrt{n_{i0}} \psi_i$, we then get 
\begin{align}
\tilde{L}  = \int d \mathbf{r}  \biggl\{ - \frac{i}{2} ( \psi_1 \dot{\psi}_1^{\ast} -  \psi_1^{\ast} \dot{\psi}_1  )
- \frac{i}{2} \tilde{\gamma} ( \psi_2 \dot{\psi}_2^{\ast} -  \psi_2^{\ast} \dot{\psi}_2  ) \nonumber \\
- [ | \nabla \psi_1|^2 + \alpha \tilde{\gamma} | \nabla \psi_2|^2
+ \frac{c_1}{2} (|\psi_1|^2-1)^2   \nonumber \\ 
+ \frac{c_2}{2} \tilde{\gamma} (|\psi_2|^2-1)^2
+ c_{2} \gamma (|\psi_1|^2 - 1)( |\psi_2|^2 - 1)] \biggl\} ,
\label{GPlagdimless}
\end{align}
where $\tilde{L} = L/\mu_1 n_{10} \xi^2$ and the parameters are given by 
\begin{align}
c_1 &= \frac{g_1 n_{10}}{\mu_1},  \hspace{3mm}  
c_2 =  \frac{g_2 n_{20}}{\mu_1},  \hspace{3mm}  
 \alpha = \frac{m_2}{m_1}, \nonumber \\  \tilde{\gamma} &= \frac{n_{20}}{n_{10}}, 
\hspace{5mm} \gamma = \frac{g_{12}}{g_2}. 
\label{parameterss}
\end{align} 
In these scales, the equilibrium amplitude of the wave function becomes $|\psi_1| = |\psi_2| = 1$. 
For simplicity, the parameters in the following discussion are put as 
$m_1 = m_2 =m$, $\mu_1 = \mu_2 =\mu$ and 
$g_1 = g_2 =g$, and thus $n_{10} = n_{20}$. Then, the parameters in Eq.~(\ref{parameterss}) become 
$\alpha =1$, $c_1 = c_2 = (1+\gamma)^{-1}$, and $\tilde{\gamma} = 1$. 
The free parameter $\gamma$ has a range $-1 < \gamma < 1$. 

The coupled GP equations are obtained by the
variational principle of the action with the Lagrangian Eq.~(\ref{GPlagdimless}) as 
(note that $\mu=1$ in our unit)
\begin{align}
i \frac{\partial \psi_1}{\partial t} &= \left( - \nabla^2 -1 + \frac{1}{1+\gamma} |\psi_1|^2 
+ \frac{\gamma}{1+\gamma} |\psi_2|^2\right) \psi_1,
\label{eq:GP1}\\
i \frac{\partial \psi_2}{\partial t} &= \left( - \nabla^2 -1 + \frac{1}{1+\gamma}  |\psi_2|^2 
+ \frac{\gamma}{1+\gamma} |\psi_1|^2\right) \psi_2. 
\label{eq:GP2}
\end{align}
The stationary coupled GP equations can be derived by neglecting the time dependence:
\begin{align}
\left( - \nabla^2  + \frac{1}{1+\gamma} |\psi_1(\mathbf{r})|^2 +  \frac{\gamma}{1+\gamma}  |\psi_2(\mathbf{r})|^2\right)\psi_1(\mathbf{r}) = \psi_1(\mathbf{r}),
\label{eq:GP3} \\ 
\left( - \nabla^2  +  \frac{1}{1+\gamma} |\psi_2(\mathbf{r})|^2 +  \frac{\gamma}{1+\gamma} |\psi_1(\mathbf{r})|^2\right)\psi_2(\mathbf{r}) = \psi_2(\mathbf{r}).
\label{eq:GP4} 
\end{align}

\subsection{Half-quantized vortex}\label{hqvintro}
Since there are miscible binary condensates in space where the condensate wave functions 
$\psi_i = \sqrt{\rho_i} e^{i \theta_i}$ are defined, 
two $U(1)$ symmetries are spontaneously broken in the system. 
Accordingly, the order parameter space is 
\begin{align}
 T^2 \simeq U(1)_1 \times U(1)_2 \simeq 
{U(1)_{\rm mass} \times U(1)_{\rm spin} \over {\mathbb Z}_2}.
 \label{eq:OPS}
\end{align} 
Here, each $U(1)_i$ ($i=1,2$) corresponds
to the phase rotation of $\psi_1$ or $\psi_2$, while
$U(1)_{\rm mass}$ and $U(1)_{\rm spin}$ 
correspond to the overall and relative phase rotations, defined by
\begin{align}
\Theta = (\theta_1 + \theta_2)/2, \quad 
\phi = (\theta_1 - \theta_2)/2 
\end{align}
 that transform as
\begin{align}
 U(1)_{\rm mass}:& \quad \psi_1 \to \psi_1 e^{i \alpha}, \quad 
 \psi_2 \to \psi_2 e^{i \alpha}, \quad \Theta \to \Theta + 2\alpha \nonumber \\
 U(1)_{\rm spin}:& \quad \psi_1 \to \psi_1 e^{i \beta}, \quad 
 \psi_2 \to \psi_2 e^{-i \beta} , \quad \phi \to \phi + 2 \beta 
 \label{eq:mass-spin}
\end{align}
whose currents are mass and pseudo-spin currents, respectively. 

The gradient of these angles can be regarded as a mass and a spin current, respectively. 
Since both the condensates $\psi_1$ and $\psi_2$ are unchanged under 
the ${\mathbb Z}_2$ action defined by $\alpha = \beta = \pi$ 
inside $U(1)_{\rm mass} \times U(1)_{\rm spin}$ 
in Eq.~(\ref{eq:mass-spin}), this ${\mathbb Z}_2$ has to be removed 
as the denominator of Eq.~(\ref{eq:OPS}). 
If the only $\psi_1$-component has a vortex $\psi_1 = \sqrt{\rho_1}e^{i\theta}$ 
with the polar angle $\theta$, the pseudo-spinor of the order parameter can be written as 
\begin{align}
\left(
\begin{array}{c} 
\psi_1 \\
\psi_2 
\end{array}
\right) 
= \left( 
\begin{array}{c} 
\sqrt{\rho_1} e^{i \theta} \\
\sqrt{\rho_2} 
\end{array}
\right) 
= \sqrt{\rho} e^{i \theta/2} \left( 
\begin{array}{c} 
\zeta_1 e^{i \theta/2} \\
\zeta_2 e^{-i \theta/2} 
\end{array}
\right),
\end{align}
where $\zeta_1^2+\zeta_2^2=1$. 
In terms of $U(1)_{\rm mass}$ and $U(1)_{\rm spin}$ in Eq.~(\ref{eq:mass-spin}), 
the both angles $\Theta$ and $\phi$ are rotated by $\pi$ with circling around a vortex. 
Since this vortex can be seen as having a half winding of $U(1)_{\rm mass}$, 
it is often called a {\it half-quantized} vortex (HQV).
The first prediction in the context of atomic-gas BEC was given by Leonhardt 
and Volovik \cite{Leonhardt}.
These similar structures have been predicted in thin films of superfluid $^3$He-A \cite{Mineev}, 
and observed for both exiton-polariton BECs \cite{Lagoudakis,Manni} 
and chiral $p$-wave superconductors \cite{Jang}.  

\begin{figure}[ht]
\centering
\includegraphics[width=1.0\linewidth,bb=0 0 396 496]{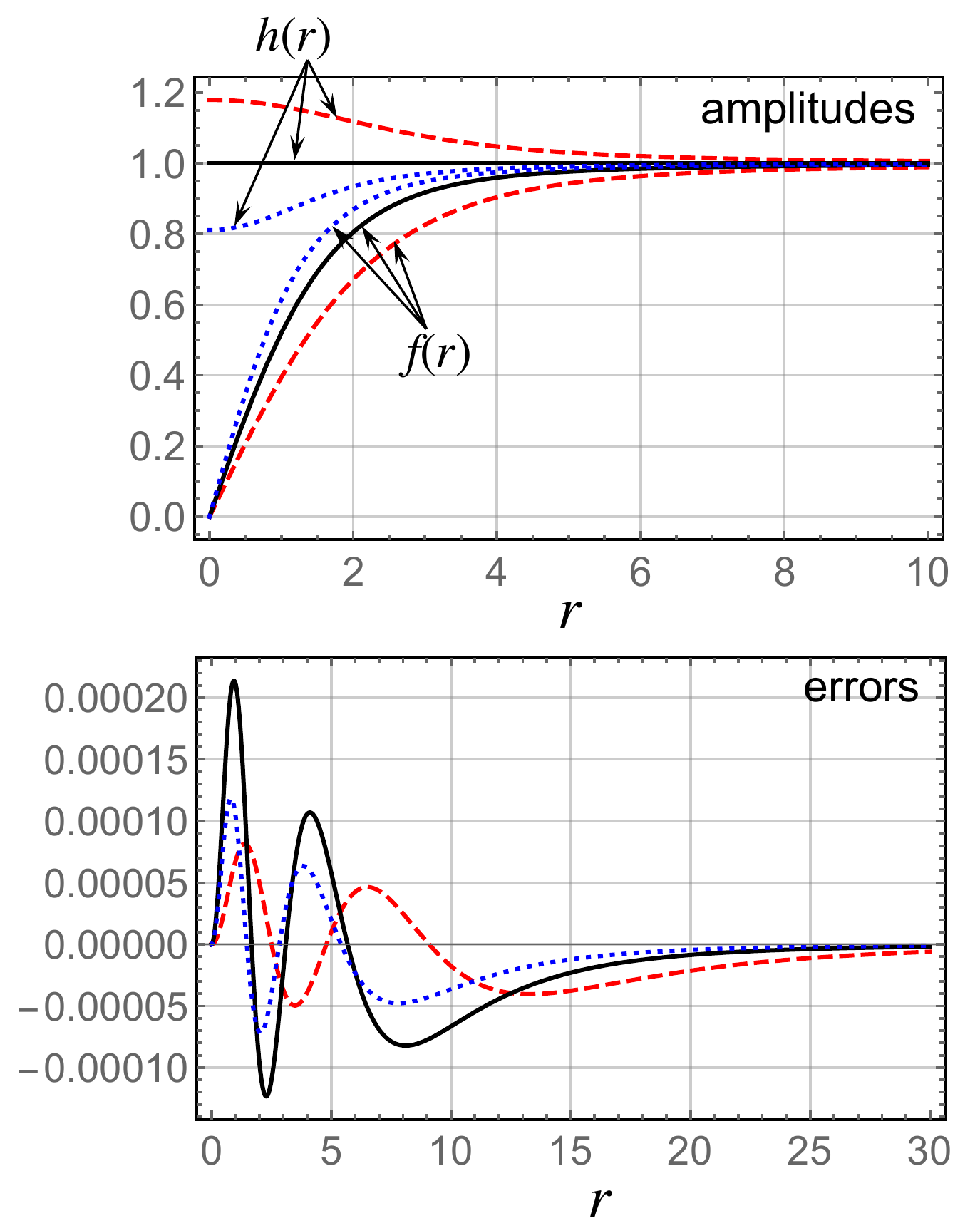} \\\
\caption{(Color online) The numerical solutions for $f(r)$ and $h(r)$  
with $\gamma=0.5$ (red long-dashed curves), $0$ (black solid curves), and $-0.5$
(blue short dashed curves) are shown in the upper panel. The lower panel
are plots of the error of the Pad\'{e} approximation from the numerical solutions: 
$\left(f_{\rm num} - f_{\text{Pad\'{e}}}\right)/f_{\rm num}$.
The style of plotting curves corresponds to the one in the upper panel.}
\label{fig:pade}
\end{figure}
The structure of a simple HQV located on the origin of the space 
along the $z$-axis can be obtained by solving Eqs.~(\ref{eq:GP3}) and (\ref{eq:GP4}) 
with an ansatz for an axially symmetric configuration 
\begin{align}
\psi_1 = \sqrt{f(r)}  e^{i\theta} , \quad
\psi_2 = \sqrt{h(r)}  ,
\label{eq:ansatz}
\end{align}
where $(r,\theta)$ is the polar coordinate and we assume that 
$\psi_1$ has a vortex with the winding number 1 and $\psi_2$ is vortex-free. 
Typical numerical solutions are shown in Fig.~\ref{fig:pade}. 
A universal feature of configuration is that the profile function $h$ of unwinding field 
at the vortex center is concave for $g_{12}<0$ 
and convex for the $g_{12} > 0$. 
This can be understood from the atom-atom interaction $g_{12}$; 
in the presence of the vortex profile for $\psi_1$ as a background, 
$\psi_2$ feels the potential $g_{12}|\psi_1|^2$
and it tends to be trapped in the vortex center for 
the repulsive interaction $g_{12}>0$ 
and to be exclusive from the vortex center for the 
attractive interaction $g_{12}<0$. 
When $\psi_2$ has a vortex and $\psi_1$ is vortex free, the role of $f$ and $h$ 
in Eq.~(\ref{eq:ansatz}) is exchanged. 

\begin{table*}
\caption{\label{tablepade} List of the values of the parameters in Eqs.~(\ref{eq:gp_varia1}) 
and (\ref{eq:gp_varia2}) as a function of $\gamma$. }
\begin{ruledtabular}
\begin{tabular}{cccccccc}
$\gamma$ &  $a_1$ & $a_2$ & $a_3$ & $b_1$ & $b_2$ & $a_0$ & $b_0$  \\
\hline
-0.9 & 0.258405 & 1.001440 & 0.128055 & 0.677999 & 0.231094 & 0.893941 & 0.302519 \\
-0.8 & 0.288090 & 0.887745 & 0.043337 & 0.632548 & 0.207753 & 0.746991 & 0.415541 \\
-0.7 & 0.328417 & 0.845890 & 0.034572 & 0.604638 & 0.185217 & 0.658306 & 0.505052 \\
-0.6 & 0.331510 & 0.810535 & 0.044151 & 0.583411 & 0.164030 & 0.592318 & 0.583911 \\
-0.5 & 0.320207 & 0.778227 & 0.054746 & 0.565440 & 0.143665 & 0.538307 & 0.657000 \\
-0.4 & 0.301442 & 0.746212 & 0.059707 & 0.549184 & 0.123534 & 0.491584 & 0.726859 \\
-0.3 & 0.279456 & 0.714102 & 0.059945 & 0.533788 & 0.102964 & 0.449661 & 0.795066 \\
-0.2 & 0.256806 & 0.682365 & 0.057508 & 0.518693 & 0.080933 & 0.411045 & 0.862748 \\
-0.1 & 0.234548 & 0.651120 & 0.053656 & 0.503478 & 0.055062 & 0.374758 & 0.930797 \\
0.0 & 0.212942 & 0.620126 & 0.049024 & --- & --- & 0.340110 & --- \\
0.1 & 0.191892 & 0.588923 & 0.043910 & 0.050700 & 0.468493 & 0.306585 & 1.071106 \\
0.2 & 0.171146 & 0.556895 & 0.038440 & 0.068459 & 0.448235 & 0.273774 & 1.144887 \\
0.3 & 0.150382 & 0.523257 & 0.032653 & 0.079615 & 0.426540 & 0.241333 & 1.222194 \\ 
0.4 & 0.129252 & 0.487011 & 0.026562 & 0.086596 & 0.402814 & 0.208962 & 1.304013 \\
0.5 & 0.107461 & 0.446909 & 0.020237 & 0.090117 & 0.376227 & 0.176376 & 1.391547 \\
0.6 & 0.084925 & 0.401450 & 0.013950 & 0.090202 & 0.345525 & 0.143300 & 1.486338 \\
0.7 & 0.061980 & 0.348770 & 0.008304 & 0.086311 & 0.308600 & 0.109448 & 1.590464 \\ 
0.8 & 0.039332 & 0.285410 & 0.003955 & 0.077039 & 0.261349 & 0.074514 & 1.706916 \\
0.9 & 0.017902 & 0.201664 & 0.000913 & 0.058583 & 0.193295 & 0.038159 & 1.840424 
\end{tabular}
\end{ruledtabular}
\end{table*}
Although the solution can be obtained only by numerically, it is practical to 
use the approximate analytical form of the profile $f$ and $h$ for calculating 
the physical quantities such as the intervortex potential. 
Their asymptotic form for large $r$ is known as \cite{Eto}
\begin{align}
f (r) \sim 1 - \frac{1}{(1-\gamma) r^2} , \hspace{4mm}
h (r) \sim  1 + \frac{\gamma}{(1-\gamma) r^2} .
\label{eq:gp_asym}
\end{align}
The miscible condition of the ground state ensures that 
the denominator is alway positive for each of Eqs.~(\ref{eq:gp_asym}), 
while $h(r)$ changes its sign 
when $\gamma < 0$. 
Recently, a variational ansatz of a HQV that describes well around its vortex core 
was considered by Mason \cite{Mason}, but it does not reproduce 
the asymptotic form for large $r$. 
Here, we apply the Pad\'{e} approximation to the profile function as
\begin{align}
f (r) &= \frac{a_0 r^2 + a_1^2 r^4 +a_3^2 (1-\gamma) r^6}{1 + a_2^2 r^2 +(a_1^2+a_3^2) r^4 + a_3^2(1-\gamma) r^6}, \label{eq:gp_varia1} \\
h (r) &= 
\left\{
\begin{array}{l}
\dfrac{b_0 + (b_1^2 + b_2^2) r^2 + b_1^2 [(1-\gamma)/\gamma] r^4}{1 + b_2^2 r^2 + b_1^2 [(1-\gamma)/\gamma] r^4} \hspace{3mm} \mathrm{for} \hspace{2mm} \gamma>0, \\ 
\dfrac{b_0 + b_1^2 r^2 + b_2^2 [(\gamma-1)/\gamma] r^4}{1 + (b_1^2+b_2^2) r^2 + b_2^2 [(\gamma-1)/\gamma] r^4}  \hspace{3mm} \mathrm{for} \hspace{2mm} \gamma<0, \label{eq:gp_varia2}
\end{array} 
\right.
\end{align}
where these functional forms are determined so as to reproduce the 
asymptotic form of Eq.~(\ref{eq:gp_asym}) for large $r$. 
The parameters in Eqs.~(\ref{eq:gp_varia1}) and (\ref{eq:gp_varia2}) are 
determined by fitting them to the numerical solution, especially 
the behavior of the solution starting from $r=0$. 
The values are summarized in Table~\ref{tablepade}. 
The error from the numerical solutions is within the range $\pm0.02\%$ as shown in Fig.~\ref{fig:pade}
For $\gamma=0$, Eq.~(\ref{eq:gp_varia2}) reduces to $h=1$. 

In the following, we denote the vortex state as $(q_1,q_2)$, where a vortex with 
the winding number $q_1$ $(q_2)$ in $\psi_1$ $(\psi_2)$-component at a certain 
position $\mathbf{r}$. Since we consider two vortices located in different positions, 
we will use the notation, e.g., $(q_1,0)$-$(0,q_2)$, where one vortex 
with $q_1$-winding in $\psi_1$ is positioned at $\mathbf{r}_{1}$ and the other 
with $q_2$-winding in $\psi_2$ is positioned at $\mathbf{r}_{2}$. 
We restrict ourselves to a single circulation $q_i = \pm1$. 
The $(1,0)$-$(0,\pm 1)$ state represents that one of two vortices is put on the different positions 
in either of the two components, and the $(1,0)$-$(\pm 1,0)$ state represents that two vortices 
are put on the different positions in one of the two components.

\section{Vortex dynamics for $(1,0)$ and $(0,\pm1)$ }\label{case2}

\subsection{Intervortex interaction}
First, we look back to the properties of intervortex interaction between two HQVs. 
The asymptotic form of the intervortex interaction for well separated vortices 
has been calculated by Eto \textit{et al}. \cite{Eto}. 
In this work, we include the short range behavior of the intervortex potential 
based on the approximation form of Eqs.~(\ref{eq:gp_varia1}) and (\ref{eq:gp_varia2}). 
We assume that the wave function with two vortices can be described by simply multiplying 
the two single vortex profiles at different positions, refereed to as the Abrikosov ansatz, as 
\begin{align}
\psi_1 & = \sqrt{\rho_1} e^{i\theta_1} = \sqrt{f(\mathbf{r} - \mathbf{r}_1) h(\mathbf{r} - \mathbf{r}_2)}
e^{i q_1 \theta^0 (\mathbf{r} - \mathbf{r}_1)}, \nonumber \\
\psi_2 & = \sqrt{\rho_2} e^{i\theta_2} = \sqrt{h(\mathbf{r} - \mathbf{r}_1) f(\mathbf{r} - \mathbf{r}_2)}
e^{i q_2 \theta^0 (\mathbf{r} - \mathbf{r}_2)} \label{initialhqv}
\end{align}
with the two phase singularities at the point $\mathbf{r}_1=(R,0)$ and 
$\mathbf{r}_2=(-R,0)$ as 
\begin{equation}
\theta^{0}(x\pm R, y) = \arctan \frac{x \pm R}{y}. \label{vorprophase}
\end{equation} 
Inserting the ansatz Eqs.~(\ref{initialhqv}) to the total energy 
\begin{align}
E_0^{(q_1,q_2)} & = \int d \mathbf{r} \biggl\{ \sum_{i=1,2}\biggr[ \frac{(\nabla \rho_i)^2}{4\rho_i} + \rho_i (\nabla \theta_i)^2 +\frac{(\rho_i - 1)^2}{2(1+\gamma)}  \biggr]  \nonumber \\ 
& + \frac{\gamma}{1+\gamma} (\rho_1 -1) (\rho_2 -1) \biggr\}, \label{dimlessanaene0}
\end{align}
we can calculate the intervortex potential from the relation \cite{Eto}
\begin{equation}
V_{12}(r_{12}) = E_0^{(1,1)}-E_0^{(1,0)}-E_0^{(0,1)}+E_0^{(0,0)},
\label{etopotori}
\end{equation}
where $r_{12} = |\mathbf{r}_2 - \mathbf{r}_1| = 2R$ 
and Eq.~(\ref{dimlessanaene0}) means $E_0^{(0,0)} = 0$. 

\begin{figure}[ht]
\centering
\includegraphics[width=1.0\linewidth,bb=0 0 524 340]{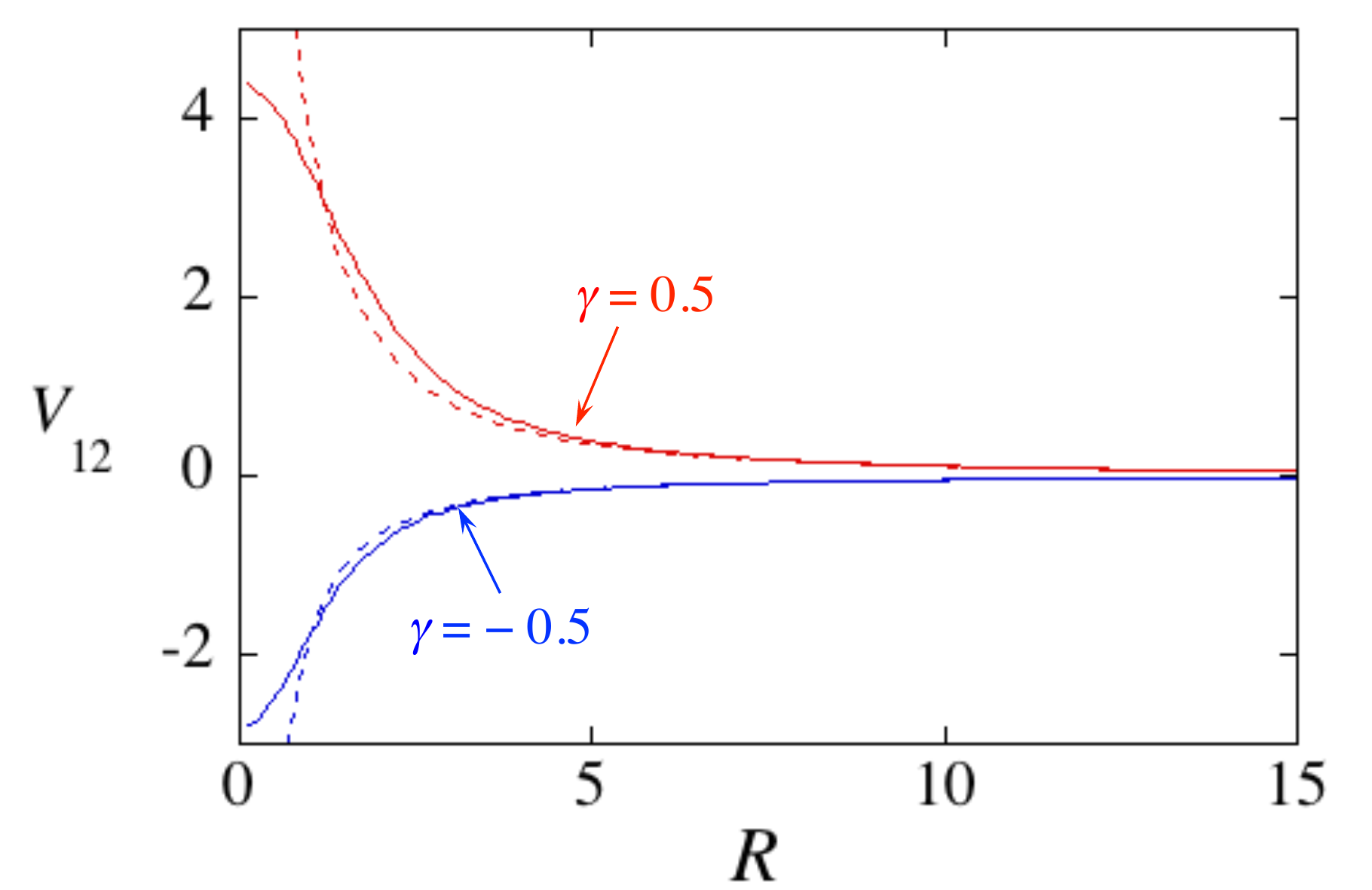} \\\
\caption{(Color online) The intervortex potential Eq.~(\ref{etopotori}) as a function of $R$ 
for $\gamma=0.5$ and $\gamma=-0.5$. The solid curve represents the 
results by numerical integration of Eq.~(\ref{dimlessanaene0}) with the 
approximated vortex profile Eqs.~(\ref{eq:gp_varia1}) and (\ref{eq:gp_varia2}), 
while the dashed curve represents the asymptotic form of Eq.~(\ref{etopot}) with 
$\xi_{\mathrm{cut}} = 0.30$ for $\gamma=0.5$ 
and $\xi_{\mathrm{cut}} = 0.18$ for $\gamma=-0.5$. 
}
\label{type12vorpot}
\end{figure}

For a certain value of $\gamma$, we calculate Eq.~(\ref{dimlessanaene0}
using the profile function Eqs.~(\ref{eq:gp_varia1}) and (\ref{eq:gp_varia2}) with the parameters 
displayed in Table~\ref{tablepade}. 
The integration was done numerically because 
the direct integration by hand yields complicated form according 
to Eqs.~(\ref{eq:gp_varia1}) and (\ref{eq:gp_varia2}).
The plot of $V_{12}$ for $\gamma=0.5$ and $-0.5$ is 
shown in Fig.~\ref{type12vorpot}. 
The vortex interaction is repulsive and attractive for $\gamma>0$ and 
$\gamma < 0$, respectively. 
For large $R$, the potential behaves as the asymptotic form 
derived by Ref.~\cite{Eto}:
\begin{equation}
V_{12}^{\mathrm{asy}} (r_{12}) = \frac{4 \pi \gamma}{1-\gamma} \frac{\log (r_{12}/2\xi_{\mathrm{cut}})}{r_{12}^2}. 
\label{etopot}
\end{equation}
Here, $\xi_{\mathrm{cut}}$ is an uncertain cut-off length, which is determined by the 
fitting to the result by numerical integration of Eq.~(\ref{etopotori}) for large $R$ \cite{Eto}; 
the values for $\gamma=\pm 0.5$ are written in the caption of Fig.~\ref{type12vorpot}. 
The numerical result matches the 
asymptotic form even for a short distance (down to $R \simeq$ 5). 
Especially, the potential for negative $\gamma$ shows a good agreement. 
Thus, we can say from this analysis that the asymptotic form of Eq.~(\ref{etopot}) 
describes $V_{12}$ well even for a short distance between two vortices. 
The deviation becomes significant for $\gamma \to 1$, because the 
vortex core extends over the space so that the description of the asymptotic form 
becomes worse. 

\subsection{Numerical simulations}\label{numety2}
Here, we study the dynamics of two interacting vortices in two-component BECs 
by numerically solving the time-dependent GP equations (\ref{eq:GP1}) and (\ref{eq:GP2}).
The initial state is prepared as follows. 
We first put the vortices using the Abrikosov ansatz with the approximated profile 
of Eqs.~(\ref{eq:gp_varia1}) and (\ref{eq:gp_varia2}) as well as 
the phase profile Eq.~(\ref{vorprophase}) at $x = + R$ in the $\psi_1$-component and 
at $x=-R$ in the $\psi_2$-component. 
Using this wave function, we obtain the more accurate initial configuration through 
the short imaginary time evolution of Eqs.~(\ref{eq:GP1}) and (\ref{eq:GP2}). 
This procedure can remove some numerical error associated with the approximation 
function Eqs.~(\ref{eq:gp_varia1}) and (\ref{eq:gp_varia2}), which improve the 
accuracy of the following real time simulations. 
During this imaginary time evolution, the position of the phase singularities 
moves but its displacement is very small. 
We impose the Neumann condition to the numerical boundary. 
The size of the numerical box is $[-40,+40]$ in unit of $\xi$ and the grid number 
is $512 \times 512$ \cite{tyuu}.  

\subsubsection{The same circulation: $(1,0)$-$(0,1)$ vortex}
\begin{figure}[ht]
\centering
\includegraphics[width=0.9\linewidth,bb=0 0 425 720]{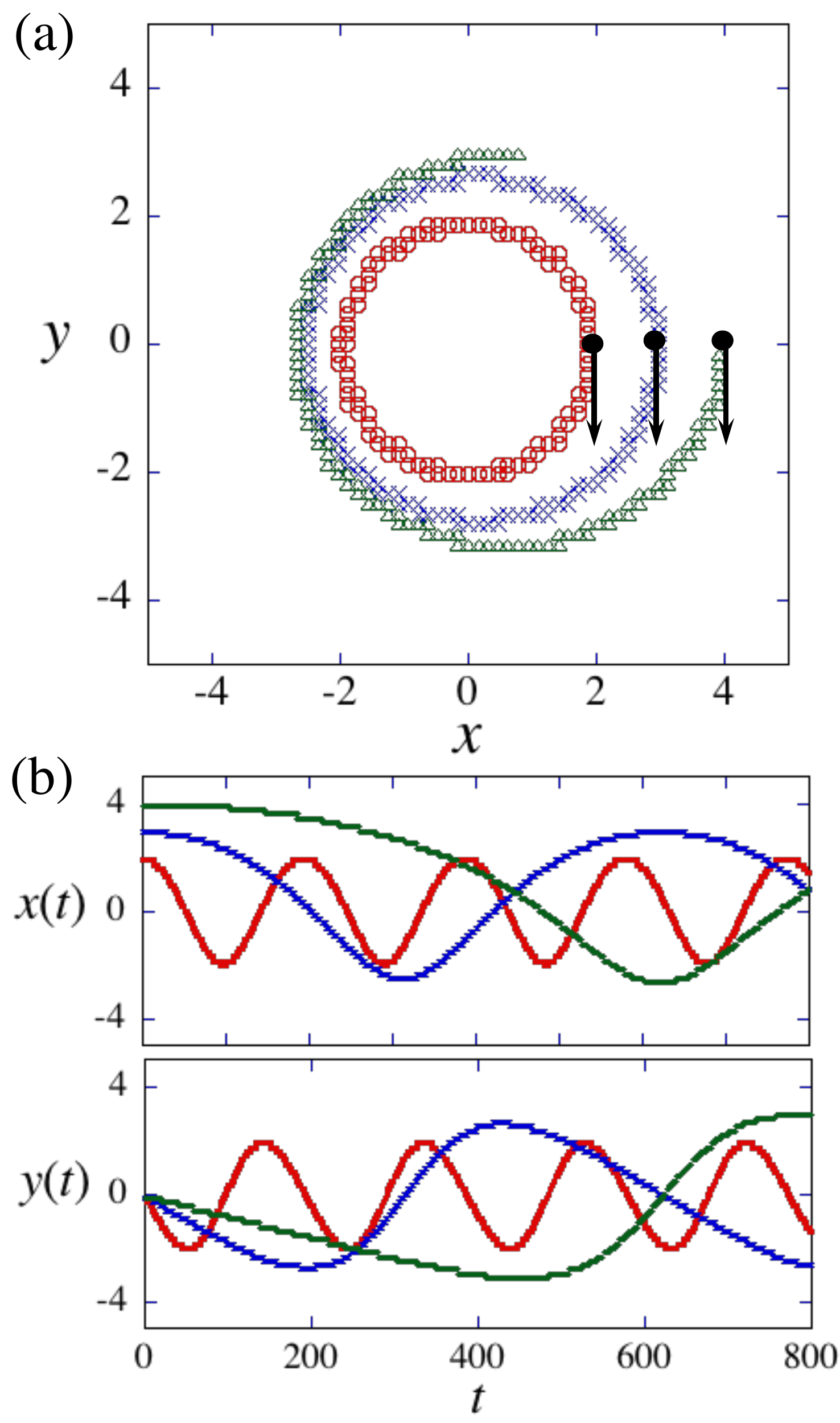} \\\
\caption{The vortex motion of two HQVs with the same circulation for $\gamma = -0.5$, 
where one of the two vortices are put on either of the two components, i.e., $(1,0)$-$(0,1)$.  
The panels (a) shows the vortex trajectories for $\psi_1$-component for the time $[0,800]$. 
The (red) circles, (blue) cross, and (green) triangles 
represents the results for initial separation $R=2$, 3, and 4. 
The initial position and the direction of the motion are indicated by black dots and allows, respectively. 
The vortex trajectory of the $\psi_2$-component is symmetric with respect to the origin.
The bottom panels (b) represent the time evolution of the vortex coordinate $x(t)$ (top) 
and $y(t)$ (bottom) in the $\psi_1$-component.}
\label{a12=-0.5}
\end{figure}

\begin{figure}[ht]
\centering
\includegraphics[width=0.9\linewidth,bb=0 0 425 720]{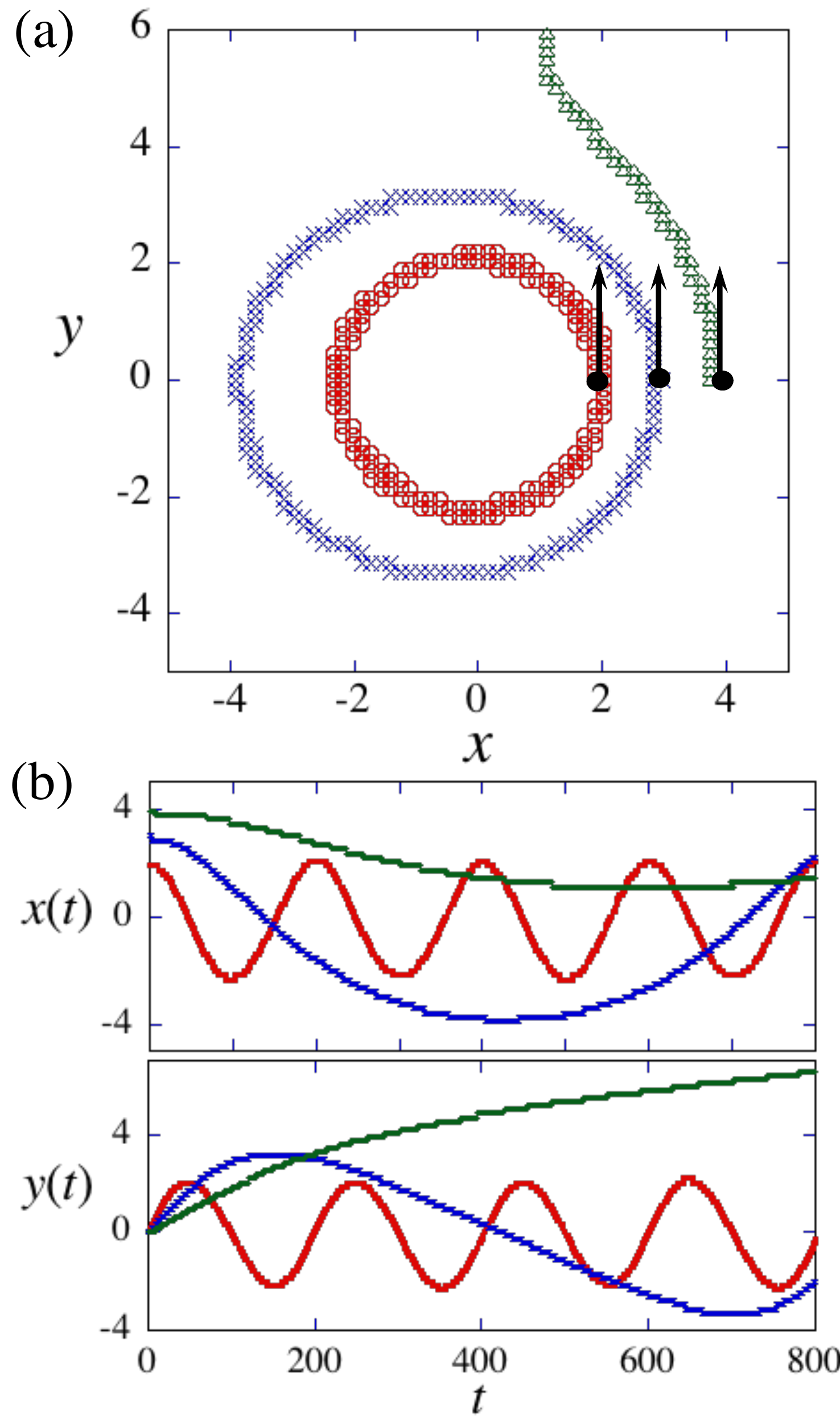} \\\
\caption{
The vortex motion of two HQVs with the same circulation 
(1,0)-(0,1) for $\gamma = 0.5$.
The plots are similar to Fig.~\ref{a12=-0.5}.
}
\label{a12=0.5}
\end{figure}

In this subsection, we focus on the dynamics of the HQVs with the same circulation, 
namely $(1,0)$-$(0,1)$. The typical vortex trajectories in the simulation is 
shown in Figs.~\ref{a12=-0.5} and \ref{a12=0.5}; only the trajectory of the vortex 
in the $\psi_1$-component is shown because that of the $\psi_2$-component is symmetric 
with respect to the origin.
For $\gamma<0$, each vortex makes a closed quasi-circular trajectory as shown in Fig.~\ref{a12=-0.5}. 
With increasing the separation $R$, the velocity of the motion becomes longer
and the trajectory is gradually distorted from the circular shape. 
The distortion occurs in such a way that the radius of the circular motion is reduced. 
We see that these behaviors are qualitatively similar to the cases with different values 
of $\gamma$, but there are some quantitative differences, e.g., the velocity of motion 
becomes slightly faster as $|\gamma|$ is increased.

For $\gamma>0$, each vortex also makes a quasi-circular trajectory as shown in Fig.~\ref{a12=0.5}, 
but the initial velocity is inverted to that in the $\gamma<0$ cases. 
In this case, the trajectory is also distorted with increasing $R$ in such a way that 
the radius of the circular motion is slightly expanded. 
For the initial separation larger than $R \simeq 4$, the vortices go away by getting off the closed trajectory. 
The velocity of motion is also monotonically decreased (increased) with $R$ ($\gamma$). 

\subsubsection{The different circulation: $(1,0)$-$(0,-1)$ vortex}   \label{case(ii)B}
Next, we study the dynamics of two HQVs with oppsite circulations $(1,0)$ and $(0,-1)$. 
It is known that an usual vortex--anti-vortex pair (vortex dipole) 
in a single-component BEC goes straightforwardly 
by keeping the distance between a vortex and an anti-vortex. 
However, the dynamics of a pair of a HQV and an anti-HQV is remarkably different. 

\begin{figure}[ht]
\centering
\includegraphics[width=0.9\linewidth,bb=0 0 425 765]{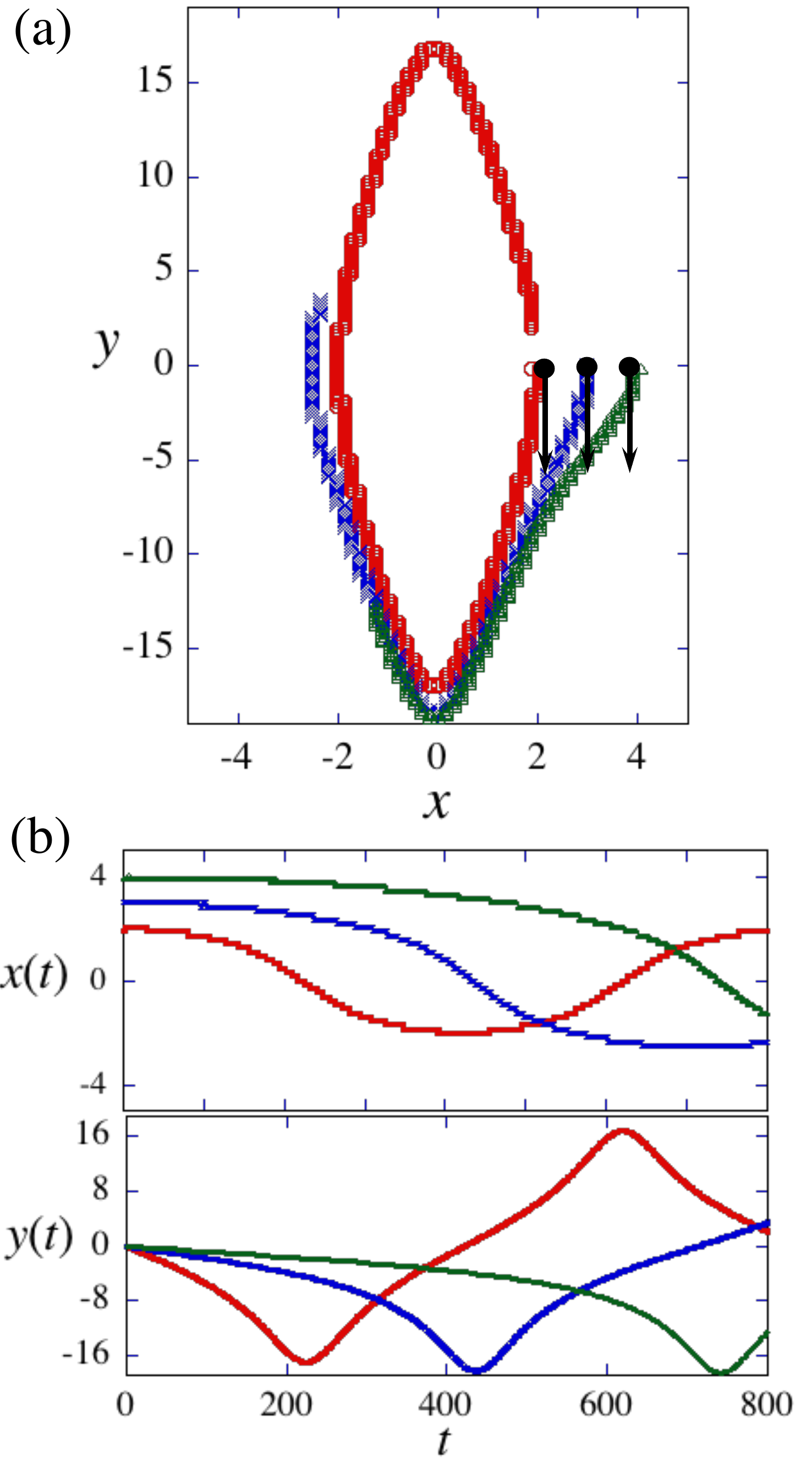} \\\
\caption{The vortex motion of two HQVs with $(1,0)$-$(0,-1)$ for $\gamma = -0.5$, 
where the initial vortex separation is $R=2$, 3, 4. 
Since the vortex trajectories in the two components are symmetric with respect 
to the $y$-axis, we only show those in the $\psi_1$-component in (a). 
The notation is similar with Fig.~\ref{a12=-0.5}. }
\label{a12=-0.5m}
\end{figure}

Figure~\ref{a12=-0.5m} shows the trajectories for $\gamma<0$. 
Since the vortex trajectories in the two components are symmetric with respect 
to the $y$-axis, we only show those in the $\psi_1$-component. 
The vortices in each component initially move to the same negative $y$-direction, 
but their separation gradually decreases. 
The vortices eventually touch each other but they are not annihilated as known 
in a pair annihilation of a vortex dipole in a single-component BEC. 
After that, they exchange their positions, 
go back to the positive $y$-direction and touch each other again. 
This motion continues and forms a closed trajectory. 
The velocity of the vortex motion becomes slower with increasing the 
initial separation $R$. 
We can see that the trajectories for $R=3$ and 4 approach to that for $R=2$, 
which implies the existence of a stable trajectory around $R=2$. 

The results for $\gamma>0$ are shown in Fig.~\ref{a1205m}. 
Although both vortices initially move to the positive $y$-direction linearly, 
they are gradually apart from each other and eventually go back to the 
negative $y$-direction. 
As $R$ increases, the initial velocity of motion becomes slower 
and the deviation from the linear trajectory takes place soon. 
\begin{figure}[ht]
\centering
\includegraphics[width=0.9\linewidth,bb=0 0 425 720]{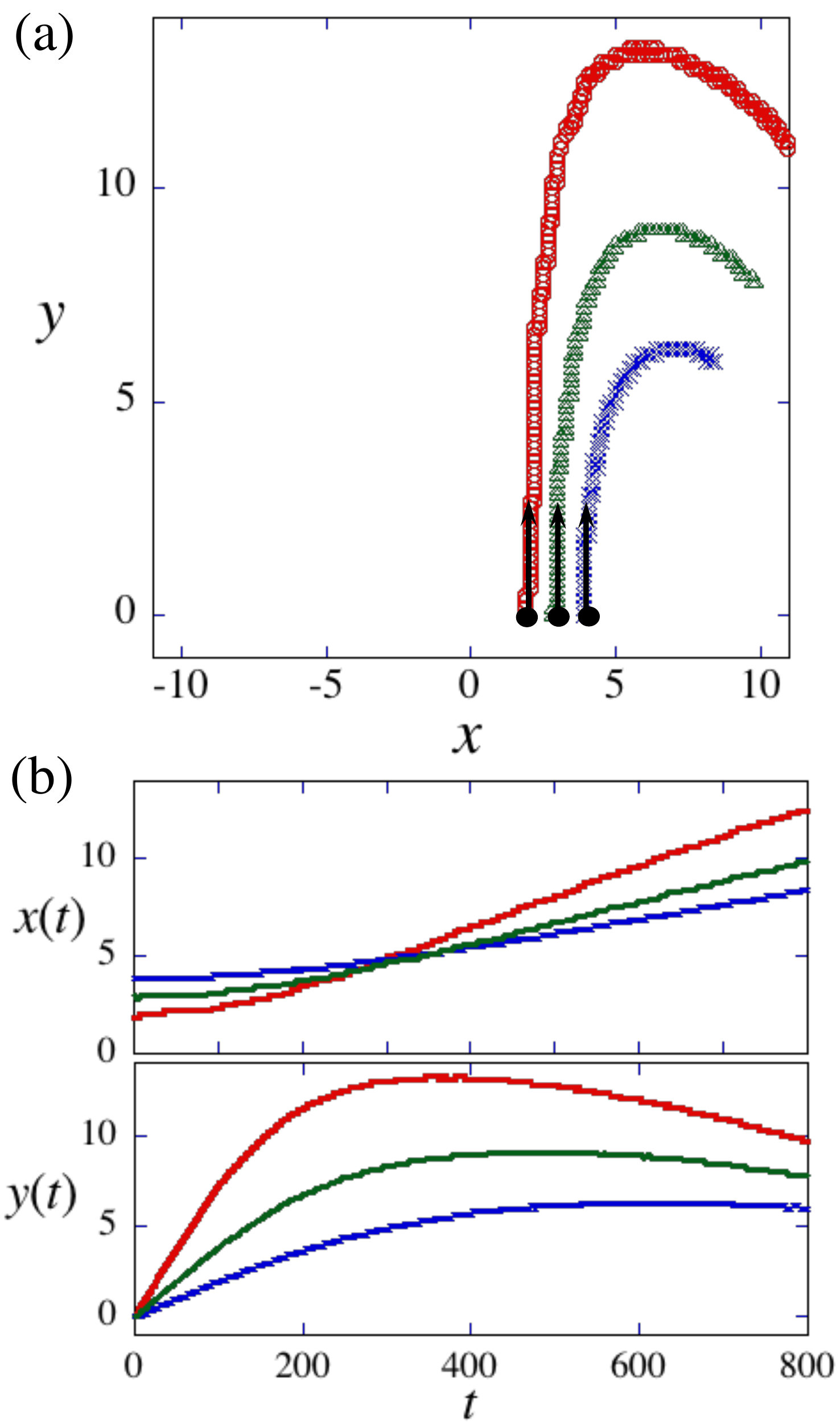} \\\
\caption{The plots similar to Fig. \ref{a12=-0.5m} for $\gamma = 0.5$. }
\label{a12=0.5m}
\end{figure}

\subsection{Point vortex approximation}\label{vpmana}
To understand the physical origin observed in the numerical simulations in Sec.~\ref{numety2}, 
we consider the dynamics of two HQVs using a point vortex approximation. 
This approximation is valid when the vortex core size is small so that the vortex can be 
seen as a point. To our knowledge, use of this model is not established 
for the dynamics of HQVs in two-component BEC. 
As shown below, our point vortex approximation can provide 
correct results only to the early stage of the dynamics, so that we need additional 
treatment to understand the full dynamics. 

We give the Abrikosov ansatz and the adiabatic approximation 
for the wave function of the $i$-th component 
\begin{align}
\psi_i(\mathbf{r},t) = \sqrt{\rho_i(\mathbf{r}, \mathbf{r}_1(t), \mathbf{r}_2(t))} 
e^{i \theta_i (\mathbf{r}, t, \mathbf{r}_i(t))},  \label{anaanz1}
\end{align} 
where 
\begin{align}
\rho_1 (\mathbf{r}, \mathbf{r}_1(t), \mathbf{r}_2(t)) 
&= f(\mathbf{r}-\mathbf{r}_{1}(t)) h(\mathbf{r}-\mathbf{r}_{2}(t)), \label{anaanz2} \\
\theta_1 (\mathbf{r}, t, \mathbf{r}_1(t)) 
&= q_1 \theta^0 (\mathbf{r}-\mathbf{r}_1(t)),  \label{anaanz3} \\
\rho_2 (\mathbf{r}, \mathbf{r}_1(t), \mathbf{r}_2(t)) 
&= h(\mathbf{r}-\mathbf{r}_{1}(t)) f(\mathbf{r}-\mathbf{r}_{2}(t)), \label{anaanz4} \\
\theta_2 (\mathbf{r}, t, \mathbf{r}_2(t)) 
&= q_2 \theta^0 (\mathbf{r}-\mathbf{r}_2(t)). \label{anaanz5}
\end{align}
Here, the amplitudes $f$ and $h$ represent the core profiles of the HQV, 
approximately given by Eqs.~(\ref{eq:gp_varia1}) and (\ref{eq:gp_varia2}), 
and $\mathbf{r}_i(t) = (x_i(t),y_i(t))$ denotes the vortex position. 
The two HQVs have the same density profile because of the symmetric choice 
of our parameters.  
The above ansatz means that two vortices are well separated and each vortex keeps its shape 
as a single vortex solution even during the dynamics. 
We expect that the density fluctuations such as phonons would have a minor effect for the 
vortex dynamics and thus neglect them. 
The phase has a profile of the vortex winding 
\begin{equation}
\theta^0 (\mathbf{r}-\mathbf{r}_i(t)) = \arctan \left[ \frac{y-y_i(t)}{x-x_i(t)} \right]. \label{phase0prof}
\end{equation}
By substituting the ansatz Eqs.~(\ref{anaanz1})-(\ref{anaanz5}) to 
the Lagrangian 
\begin{align}
L &= -\int d \mathbf{r} [
\rho_1 \dot{\theta}_{1} + \rho_2 \dot{\theta}_{2} ] - E_0^{(q_1,q_2)} \label{dimlessanalag}, 
\end{align}
we can derive the Euler-Lagrange equations for the vortex coordinates 
$\mathbf{r}_i(t)$. After some calculations, the equations of motion can be written as 
\begin{align}
2 \pi q_1 \dot{y}_1 &= - \frac{\partial V_{12}}{\partial x_1} , \label{vpmotion01} \\
-2 \pi q_1 \dot{x}_1 &= - \frac{\partial V_{12}}{\partial y_1} , \label{vpmotion02} \\
2 \pi q_2 \dot{y}_2 &= - \frac{\partial V_{12}}{\partial x_2} , \label{vpmotion03} \\
-2 \pi q_2 \dot{x}_2 &= - \frac{\partial V_{12}}{\partial y_2} .\label{vpmotion04}
\end{align}
Here, $V_{12}$ represents the inter-vortex potential given by Eq.~(\ref{etopotori}).
The details of the derivation are described in the Appendix \ref{appa}. 
The equations are essentially similar to those used in a single-component 
system \cite{Middelkamp,Navarro}, where the vortex motion can be described 
by the balance between the Magnus force and the intervortex force. 

The properties of the vortex motions depend on the sign of $q_i$. 
Let us see the case for the $(1,0)$-$(0,1)$ vortex [$(q_1,q_2)=(1,1)$]. 
One can show that $\mathbf{r}_1 + \mathbf{r}_2$ is 
a constant of motion. For the symmetric initial position such as $x_{2}(0) = - x_1(0)$ 
as in our simulations, the above constants are zero and the relative distance $r_{12}$ is also 
time-independent. 
Thus, we obtain the equation of motion for $\mathbf{r}_i$ as
\begin{equation}
\ddot{\mathbf{r}}_i = - \left[ \frac{1}{\pi r_{12}} \frac{\partial V_{12}}{\partial r_{12}} \right]^2 \mathbf{r}_i. \label{unicycmo0}
\end{equation}
This represents an uniform circular motion and the rotation frequency depends on 
the $\gamma$. 
From Eq.~(\ref{vpmotion01}), the initial velocity $\dot{y}_1(0)$ is positive (negative) for $\gamma>0$ ($\gamma<0$), 
which is consistent with the numerical results. 

For the $(1,0)$-$(0,-1)$ vortex [$(q_1,q_2)=(1,-1)$], one can show that $\mathbf{r}_1 - \mathbf{r}_2 $ is a constant of motion. 
For an appropriate choice of the initial condition such as $x_1(0) - x_2(0) =  2R$ and $y_1(0) - y_2(0) = 0$ 
as in our simulations, the solution is given by 
\begin{align}
x_{1} &= R, \hspace{5mm} x_2 = -R  \nonumber \\
y_1 &= y_2=\frac{1}{2\pi} \left( \frac{\partial V_{12}}{\partial r_{12}} \right) t. \label{unilinmo0}
\end{align}
This is exactly an uniform linear motion for both vortices. 
The initial velocity $\dot{y}_1(0)$ is positive (negative) for $\gamma>0$ ($\gamma<0$), 
which is also consistent with the numerical results. 

Therefore, the vortex point model can explain partly the numerical results, but 
cannot describe some observed dynamics, namely, the deviation from the circular 
trajectory seen in Figs.~\ref{a12=-0.5} and \ref{a12=0.5} as well as the deviation 
from the linear trajectory in Figs.~\ref{a12=-0.5m} and \ref{a12=0.5m}. 
Understanding these observation would need more effects beyond the ansatz 
Eqs.~(\ref{anaanz1})-(\ref{anaanz5}). 

\begin{figure}[ht]
\centering
\includegraphics[width=1.0\linewidth,bb=0 0 453 567]{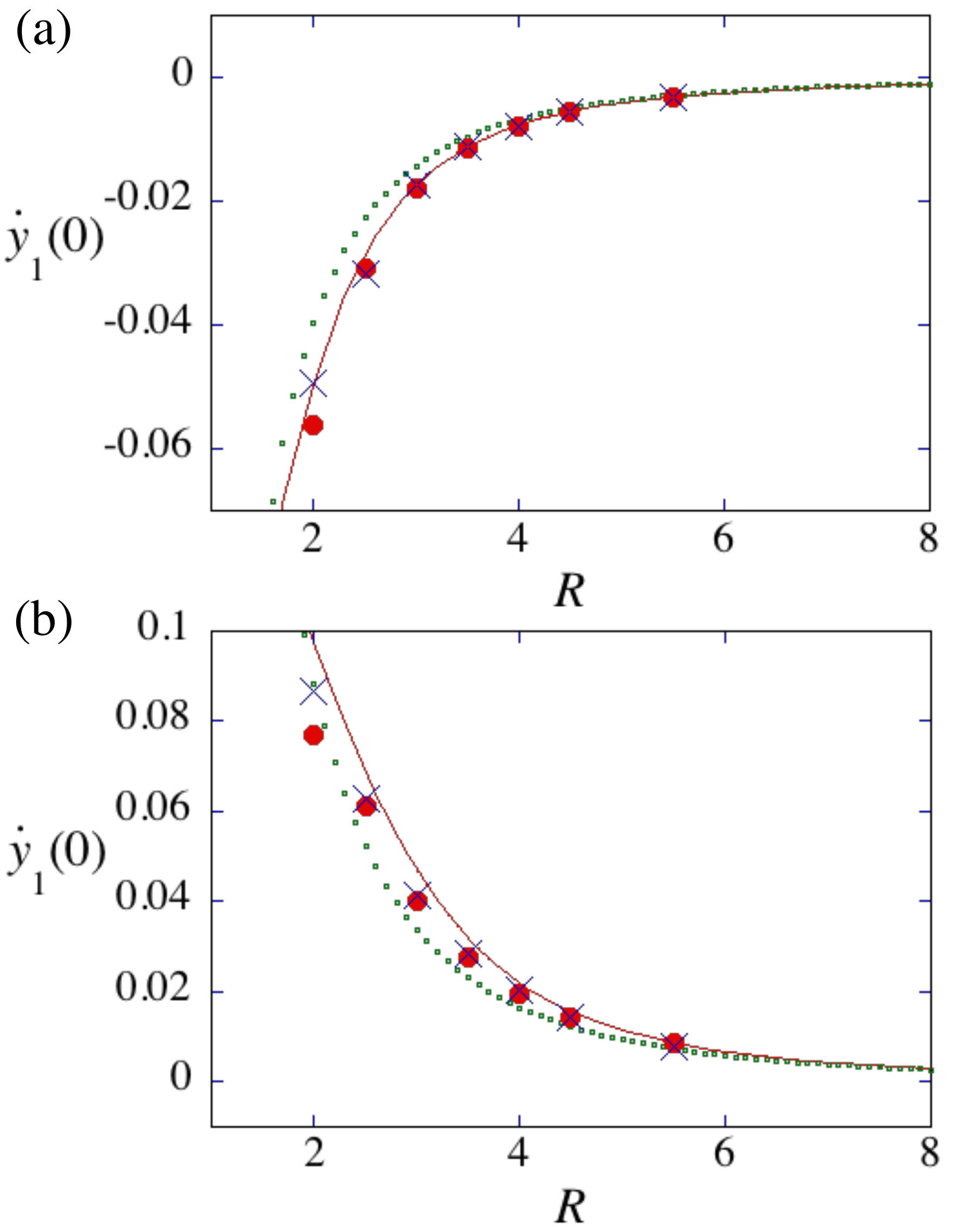} \\\
\caption{The comparison of the initial velocity of the HQV in the $\psi_1$-component between 
the numerical simulations and evaluation from the intervortex force by using the Abrikosov ansatz 
(a) $\gamma = -0.5$ and (b) $\gamma = 0.5$. 
We plot the initial velocity $\dot{y}(0)$ as a function of the initial separation $R$.
The (red) circles and (blue) crosses represent 
the numerical results of the GP equation for $(1,0)$-$(0,1)$ and $(1,0)$-$(0,-1)$, respectively. 
The initial velocity is estimated by the linear fit of the short time behavior of 
$y_1(t)$ from $t=0$.
The solid curve represents $(\partial V_{12}/\partial r_{12}) /2 \pi$, where $V_{12}$ 
is calculated from Eq.~(\ref{etopotori}) with the approximated profile function 
of Eqs.~(\ref{eq:gp_varia1}) and (\ref{eq:gp_varia2}). 
The dotted curve corresponds to the results obtained from the asymptotic intervortex interaction Eq.~(\ref{etopot}): 
$\dot{y}_1 (0) = \frac{\gamma}{q_1 (1-\gamma)} \frac{2 \ln (R/\xi_{\mathrm{cut}}) -1}{4 R^3}$ with $q_1 = +1$. }
\label{inivelocom}
\end{figure}
According to Eqs.~(\ref{vpmotion01})-(\ref{vpmotion04}), the qualitative properties of vortex motion depend 
only on the circulation of the vortices, being independent of the intervortex interaction. 
The intervortex interaction determines the initial velocities of the vortices. 
The initial direction of the vortex motion is determined by the sign of $\gamma$, 
while the magnitude of the initial velocity is dependent on the 
value of $\gamma$ irrespective of the sign of $q_i$. 
We compare the initial velocity of $y_1$ obtained from the intervortex 
force calculated from both the asymptotic potential Eq.~(\ref{etopot}) and 
the numerical integration of Eq.~(\ref{etopotori}) with those 
obtained in the numerical simulations, which is shown in Fig.~\ref{inivelocom}. 
For $R \geq 6$ the vortex motion 
becomes quite slow and it is difficult to obtain an accurate velocity from the numerical data.
The numerical results agree very well with the initial velocity 
obtained by the numerical integration of Eq.~(\ref{etopotori}) for $\gamma < 0$, 
while for $\gamma>0$ the numerical results are slightly deviated from that 
in the range $R \leq 4$. This can be explained by including the 
neglected terms when we derive Eqs.~(\ref{vpmotion01})-(\ref{vpmotion04}), 
which will be discussed in the Appendix \ref{additionalc}. 
Therefore, the initial stage of the dynamics can be captured by the vortex-point model 
Eqs.~(\ref{vpmotion01})-(\ref{vpmotion04}). The asymptotic form of the 
intervortex potential approximately works well even for a short distance between two 
vortices. 

\section{Vortex dynamics for (1,0)-($\pm 1$,0)} \label{case1}
\subsection{Intervortex potential}
Next, we consider the vortex interaction and dynamics when the two vortices are put in the 
same component, say $\psi_1$-component, along with the procedure similar to 
the previous section. 

First, we consider the intervortex potential. It is known that for well separated 
vortices at $\mathbf{r}_1 = (R,0)$ and $\mathbf{r}_2= (-R,0)$ 
in a single-component BEC the intervortex interaction is given 
by \cite{Pethickbook}
\begin{equation}
V_{12} = 2 \pi \ln \frac{R^2 + \Lambda^2}{4 R^2}. 
\label{type0potform}
\end{equation}
Here, we use the scale $\xi$ and $\tau$ to get the dimensionless form of the potential 
and $\Lambda$ is the system size.
For $(1,0)$-$(\pm 1,0)$ states of two-component BECs, the vortex 
interaction should be modified by the presence of intercomponent interaction. 
However, Eto \textit{et al}. \cite{Eto} showed that, using the asymptotic form 
Eq.~(\ref{eq:gp_asym}) of the vortex profile, the intervortex interaction is 
independent of the $\gamma$, namely $g_{12}$, being 
given by Eq.~(\ref{type0potform}) for the leading order of the vortex separation. 

\begin{figure}[ht]
\centering
\includegraphics[width=0.9\linewidth,bb=0 0 396 496]{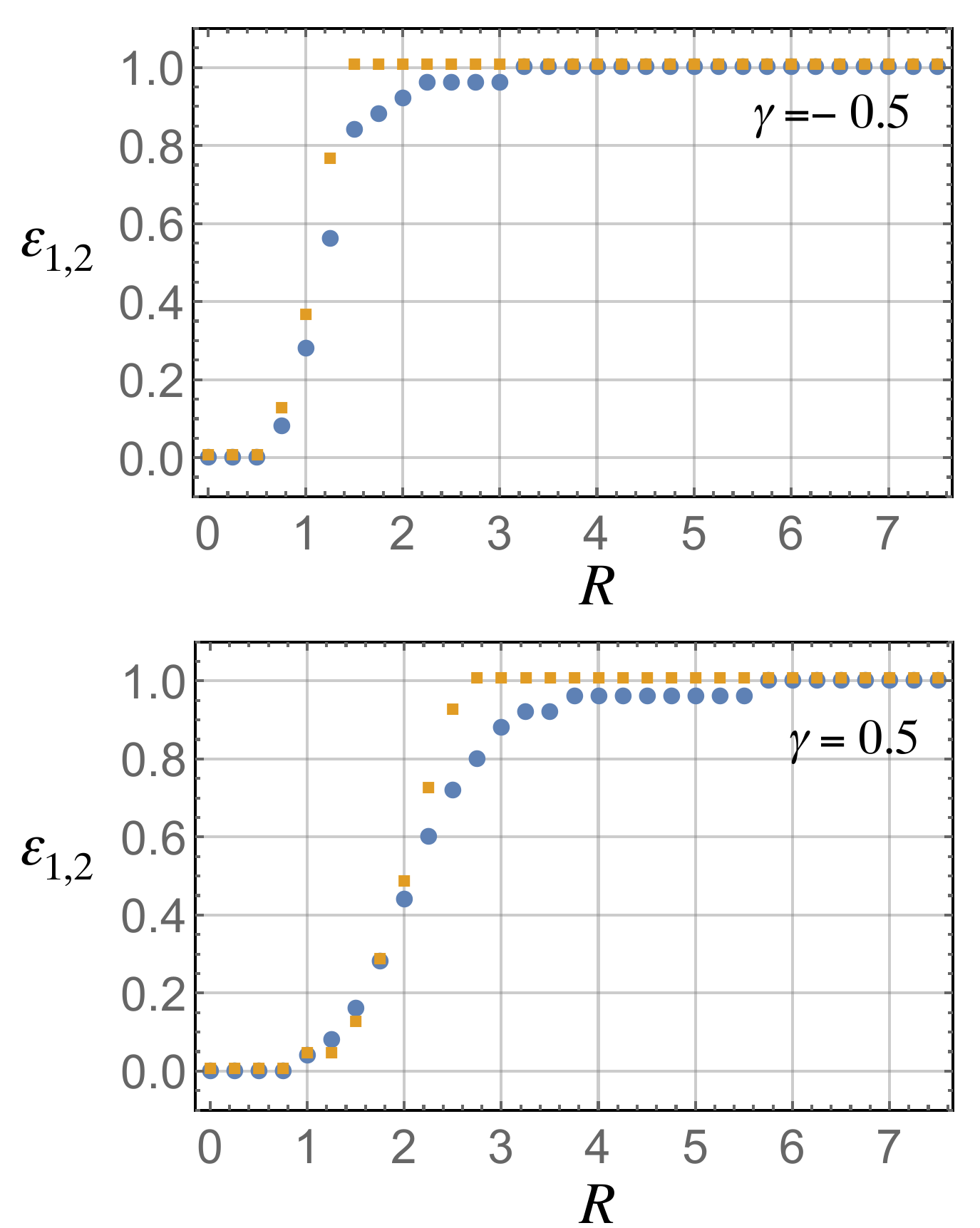} \\\
\caption{(Color online) The variational parameters $\epsilon_1$ (blue circles) and
$\epsilon_2$ (yellow squares) for $\gamma = -0.5$ (upper panel) and
$\gamma = 0.5$ (lower panel). The horizontal axes are vortex separations $R$.}
\label{fig:epsilons}
\end{figure}

We expect that the $\gamma$-dependence may appear in the short 
range property of the intervortex potential. We thus calculate the intervortex potential 
without using the asymptotic form. 
The simple Abrikosov ansatz for the wave funtion is given by
\begin{align}
\psi_1 &= \sqrt{f(\mathbf{r} - \mathbf{r}_1) f(\mathbf{r} - \mathbf{r}_2)}
e^{i q_1 \theta^0 (\mathbf{r} - \mathbf{r}_1) + i q_2 \theta^0 (\mathbf{r} - \mathbf{r}_2) }, \nonumber \\
\psi_2 &= \sqrt{h(\mathbf{r} - \mathbf{r}_1) h(\mathbf{r} - \mathbf{r}_2)} \label{initialhqv0}
\end{align}
where $f$ and $h$ are given by Eqs.~(\ref{eq:gp_varia1}) and (\ref{eq:gp_varia2}).
However, for nearly approaching vortices, the simple Abrikosov ansatz
does not  work in this situation, because two $(1,0)$-$(1,0)$ vortices must be converted to 
a vortex with double winding number in the first component $(2,0)$,  
and two $(1,0)$-$(-1,0)$ vortices must be annihilated 
for vanishing separation $R=0$. 
Thus, we have to consider a more suitable ansatz of the vortex profile to 
reproduce this situation. For $(1,0)$-$(1,0)$ state 
we can introduce the improved Abrikosov ansatz following the work \cite{Jacobs} as
\begin{widetext}
\begin{align}
\psi_1 &= (z-z_1)(z-z_2) \biggl[ \epsilon_1 \frac{\sqrt{f(\mathbf{r} - \mathbf{r}_1) f(\mathbf{r} - \mathbf{r}_2)}}
{|\mathbf{r} - \mathbf{r}_1| |\mathbf{r} - \mathbf{r}_2|}  
+(1-\epsilon_1) \frac{\sqrt{F(r)}}{r^2} \biggr], \label{impabrianz1} \\
\psi_2 &= \epsilon_2 \sqrt{h(\mathbf{r} - \mathbf{r}_1)h(\mathbf{r} - \mathbf{r}_2)} 
+ (1-\epsilon_2) \sqrt{H(r^{\ast})} \label{impabrianz2}
\end{align}
with $z=x+iy$, $\mathbf{r}_1 = (R,0)$ and $\mathbf{r}_2= (-R,0)$. Here, 
$F(\mathbf{r})$ and $H(\mathbf{r})$ represent the profile function 
of the double-quantized vortex and the corresponding unwinding field, 
respectively. 
Using the Pad\'{e} approximation, we obtain these profile 
functions as 
\begin{align}
F (r) &= \frac{A_0r^4 + A_1^2 r^6 +[A_2^2 (1-\gamma)/4] r^8}{1 + A_3^2 r^2 +A_4^2 r^4 +(A_1^2+A_2^2) r^6 + [A_2^2(1-\gamma)/4] r^8}, \label{doubleqv1} \\
H (r) &= 
\left\{
\begin{array}{l}
\dfrac{B_0 + B_2^2 r^2 +(B_1^2+B_4^2) r^4 + B_1^2 [(1-\gamma)/4\gamma] r^6}{1 + B_3^2 r^2 + B_4^2 r^4 + B_1^2 [(1-\gamma)/4\gamma] r^6} \hspace{3mm} \mathrm{for} \hspace{2mm} \gamma>0, \\ 
\dfrac{B_0 + B_2^2 r^2 + B_1^2 r^4 + B_4^2 [(\gamma-1)/4\gamma] r^6}
{1 + B_3^2 r^2 +(B_1^2+B_4^2) r^4 + B_4^2 [(\gamma-1)/4\gamma] r^6}  \hspace{3mm} \mathrm{for} \hspace{2mm} \gamma<0. \label{doubleqv2}
\end{array} 
\right.
\end{align}
\end{widetext}
As done in Sec.~\ref{hqvintro}, the parameters are determined by the fitting to 
the numerical solutions, as listed in Table~\ref{tablepade2}. 
The parameters $\epsilon_{1,2}$ represent the weight between the first and second terms, 
being determined variationally;  $\epsilon_{1,2}=1$ reproduces an asymptotic configuration 
of the separated vortices, while $\epsilon_{1,2}=0$ does the $(2,0)$ vortex. 
Thus, Eqs.~(\ref{impabrianz1}) and (\ref{impabrianz2}) nicely improve the simple
Abrikosov ansatz around zero vortex separation $R=0$. Concrete values of $\epsilon_{1,2}$ are shown
in Fig.~\ref{fig:epsilons}.

In the second term the factor $|z^2-R^2|/r^2$ has the effect of replacing the double zero 
of $F$ at the origin with two zeros at $z= \pm R$. 
Because of this change of the zero points in $\psi_1$, we also modify the 
profile of $H$ by replacing $r \to r^\ast \equiv \sqrt{|\mathbf{r} - \mathbf{r}_1| |\mathbf{r} - \mathbf{r}_2|}$. 
Using these profile functions, we substitute the ansatzs Eqs.~(\ref{impabrianz1}) 
and (\ref{impabrianz2}) to the total energy and find the 
optimal values of $\epsilon_1$ and $\epsilon_2$. 
As expected, this optimal $\epsilon_{1,2}$ are small for small $R$, 
but quickly approaches to 1 as $R$ increases past a value of $\approx 4$, 
as shown in Fig.~\ref{fig:epsilons}.
\begin{table*}
\caption{\label{tablepade2} List of the values of the parameters in Eqs.~(\ref{doubleqv1}) 
and (\ref{doubleqv2}) as a function of $\gamma$. }
\begin{ruledtabular}
\begin{tabular}{ccccccccccc}
$\gamma$ &  $A_1$ & $A_2$ & $A_3$ & $A_4$ & $B_1$ & $B_2$ & $B_3$ & $B_4$ & $A_0$ & $B_0$  \\
\hline
-0.9 & 0.000000 & 0.044085 & 0.655626 & 0.361754 & 0.174282 & 0.286248 & 0.369131 & 0.027326 & 0.123596 & 0.168104 \\
-0.8 & 0.069187 & 0.030517 & 0.628252 & 0.328574 & 0.153605 & 0.297146 & 0.359327 & 0.032409 & 0.095082 & 0.274568 \\
-0.7 & 0.086909 & 0.028872 & 0.635450 & 0.312520 & 0.139898 & 0.301351 & 0.349658 & 0.030685 & 0.078140 & 0.370791 \\
-0.6 & 0.085943 & 0.031564 & 0.626964 & 0.294624 & 0.129413 & 0.302121 & 0.339959 & 0.027272 & 0.065735 & 0.462552 \\
-0.5 & 0.075776 & 0.032859 & 0.599957 & 0.272692 & 0.120663 & 0.300841 & 0.330249 & 0.023438 & 0.055808 & 0.552190 \\
-0.4 & 0.064181 & 0.031788 & 0.567658 & 0.249943 & 0.112920 & 0.298138 & 0.320411 & 0.019539 & 0.047479 & 0.640957 \\
-0.3 & 0.054056 & 0.029371 & 0.536912 & 0.228342 & 0.105771 & 0.294312 & 0.310305 & 0.015658 & 0.040294 & 0.729641 \\
-0.2 & 0.045799 & 0.026410 & 0.509087 & 0.208299 & 0.098960 & 0.289497 & 0.299789 & 0.011749 & 0.033992 & 0.818786 \\
-0.1 & 0.034307 & 0.023979 & 0.467809 & 0.187238 & 0.092311 & 0.283730 & 0.288728 & 0.007576 & 0.028408 & 0.908799 \\
0 & 0.032810 & 0.020333 & 0.457036 & 0.171538 & --- & --- & --- & --- & 0.023439 & --- \\
0.1 & 0.000515 & 0.083379 & 0.388462 & 0.334831 & 0.006105 & 0.269140 & 0.264375 & 0.078768 & 0.019014 & 1.092646 \\
0.2 & 0.000128 & 0.120969 & 0.334556 & 0.495026 & 0.007578 & 0.260059 & 0.250736 & 0.071754 & 0.015086 & 1.186947 \\ 
0.3 & 0.000613 & 0.134327 & 0.254264 & 0.582265 & 0.007960 & 0.249488 & 0.235824 & 0.064565 & 0.011627 & 1.283069 \\
0.4 & 0.000233 & 0.086125 & 0.259538 & 0.406030 & 0.007620 & 0.237060 & 0.219329 & 0.057114 & 0.008617 & 1.381131 \\
0.5 & 0.000529 & 0.018412 & 0.283458 & 0.116434 & 0.006684 & 0.222217 & 0.200816 & 0.049309 & 0.006048 & 1.481185 \\
0.6 & 0.007453 & 0.005468 & 0.280586 & 0.070610 & 0.005168 & 0.204059 & 0.179625 & 0.041037 & 0.003919 & 1.583190 \\
0.7 & 0.007135 & 0.002747 & 0.271297 & 0.055117 & 0.002861 & 0.181002 & 0.154644 & 0.032158 & 0.002236 & 1.686947 \\
0.8 & 0.003942 & 0.001296 & 0.222135 & 0.036842 & 0.000000 & 0.152068 & 0.125377 & 0.022516 & 0.001009 & 1.791966 \\
0.9 & 0.001345 & 0.000358 & 0.154762 & 0.018222 & 0.088522 & 1.623077 & 1.178358 & 0.109412 & 0.000257 & 1.897163 
\end{tabular}
\end{ruledtabular}
\end{table*}

\begin{figure}[ht]
\centering
\includegraphics[width=0.9\linewidth,bb=0 0 540 708]{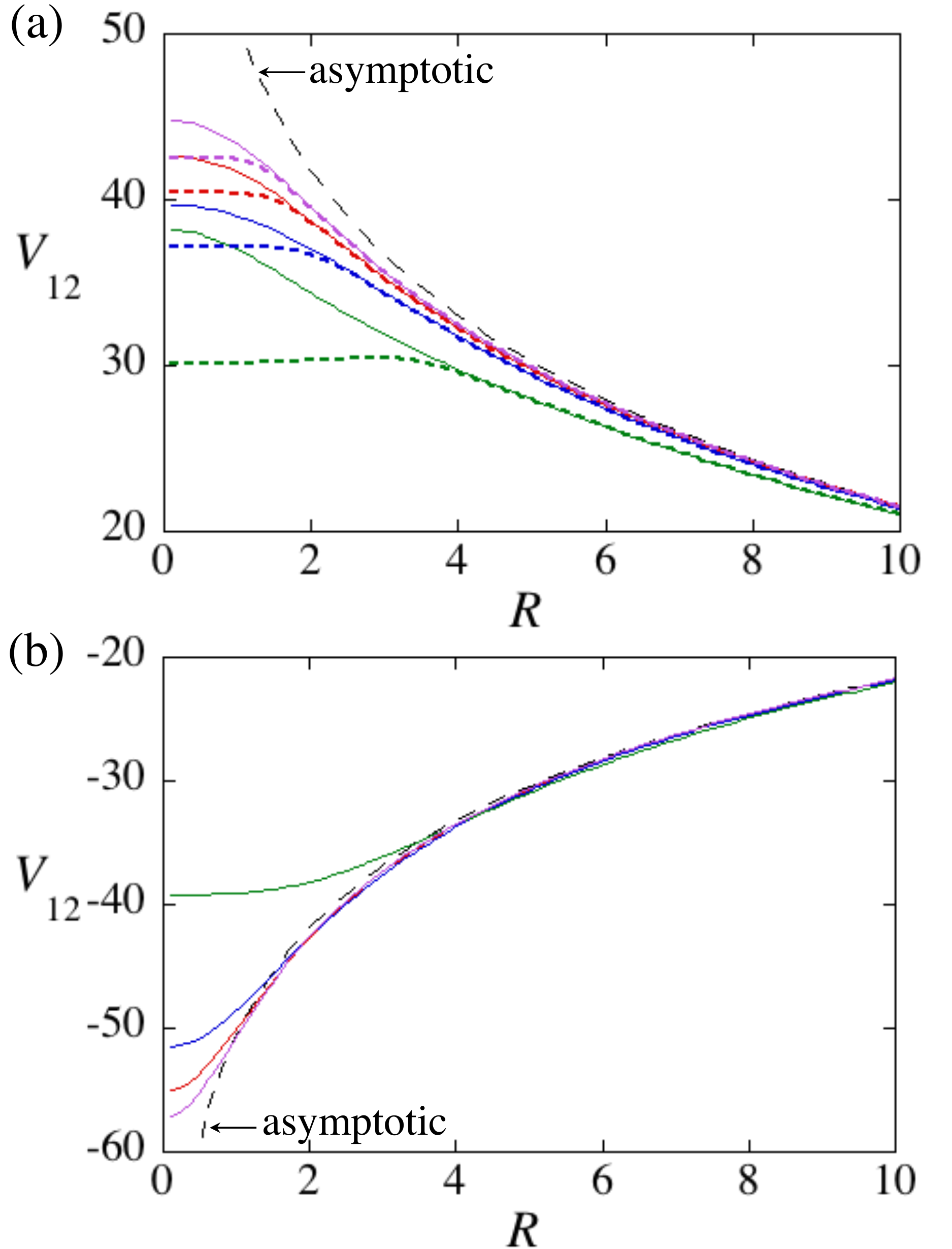} \\\
\caption{(Color online) The intervortex potential for (a) $(1,0)$-$(1,0)$ 
and (b) $(1,0)$-$(-1,0)$ case, as a function of the vortex separation $R$. 
The plots correspond to $V_{12}$ for $\gamma=-0.4$, 0, 0.4, 0.8 from 
top to bottom curves in (a) and vice versa for (b). 
In (a), the results obtained by two different methods are distinguished by the solid 
(simple Abrikosov) and the bold-dashed (improved Abrikosov) curves. 
In (b), we only show the result of the simple Abrikosov ansatz. 
The asymptotic potential Eq.~(\ref{type0potform}) is also plotted by a thin-dashed curve 
for both (a) and (b).
}
\label{type0rot}
\end{figure}
In Fig.~\ref{type0rot}, we plot the intervortex potential as a function of $R$, 
where we calculate $V_{12}$ using the simple Abrikosov ansatz Eq.~(\ref{initialhqv0}). 
For the $(1,0)$-$(1,0)$ vortex, we also calculate with the improved 
ansatz Eqs.~(\ref{impabrianz1}) and (\ref{impabrianz2}); 
we do not consider such an improvement of ansatz for the $(1,0)$-$(-1,0)$ case.  
The potential is slowly decaying function with increasing in $R$ and 
asymptotically described by Eq.~(\ref{type0potform}). 
We can confirm that the potential has a $\gamma$-dependance in a short distance. 
When we use the improve Abrikosov ansatz, the increase in $V_{12}$ for $R \to 0$ 
is strongly suppressed. A plateau appears at $R=0$, and moreover $V_{12}$ seems 
to form a local minimum at $R=0$ for $\gamma$ near $1$. However, as 
seen below, this local minimum is a consequence of the breakdown 
of the ansatz Eqs.~(\ref{impabrianz1}) and (\ref{impabrianz2}) 
around an intermediate region between the coincident vortices and well-separated vortices. 

\subsection{Numerical simulations}
Next, we consider the vortex motion by directly solving the GP equation. 
The initial state is prepared as the similar way in Sec.~\ref{numety2}. 
However, during the short imaginary time evolution, the vortices tend to move 
repulsively for the $(1,0)$-$(1,0)$ case and attractively for the $(1,0)$-$(-1,0)$ case. 
Thus, we stop the evolution when the vortices come to the expected position. 

\subsubsection{The same circulation: $(1,0)$-$(1,0)$}   
\begin{figure}[ht]
\centering
\includegraphics[width=0.9\linewidth,bb=0 0 269 453]{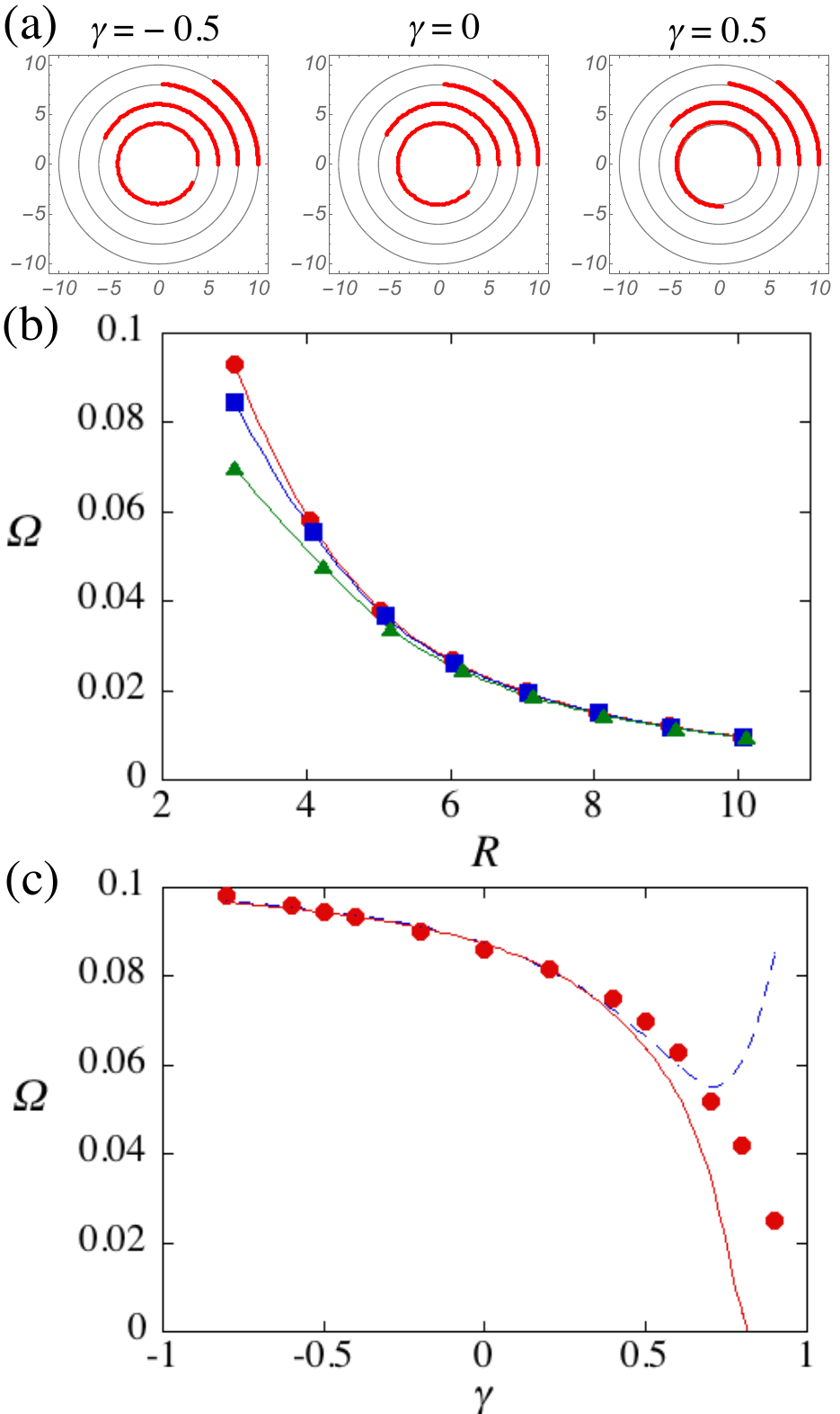} \\\
\caption{(Color online) 
Vortex dynamics of two corotating vortices in the $\psi_1$-component $(1,0)$-$(1,0)$. 
(a) Trajectories of one of the vortices initially positioned at $x=+R$ 
for $\gamma=-0.5$, 0, 0.5 from left to right. 
Circles with the radius $R=$4, 6, 8, 10 are represented by black solid curves, and 
the vortex trajectory from $t=0$ to $t=100$ are by red bold curves. 
(b) Plots of the rotation frequency $\Omega$ as a function of the 
vortex separation $R$ for $\gamma=0$ (squares), 0.5 (triangles), and $-5$ (circles). 
(c) Plots $\Omega$ as a function of $\gamma$ for $R=3$. 
The numerical results are compared with the two methods to calculate $V_{12}$: 
simple Abrikosov ansatz (dashed-curve) and improved Abrikosov ansatz (solid-curve). 
}
\label{type0freq}
\end{figure}

For $\gamma = 0$, the dynamics must be equivalent to that of a single component BECs, 
because the two components are completely decoupled. 
Namely, for vortices with the same circulation $(1,0)$-$(1,0)$ or $(-1,0)$-$(-1,0)$, 
the two vortices make a circular motion. This is actually reproduced in the simulation. 
With increasing the initial separation $R$, the rotation frequency becomes smaller as shown in 
Fig.~\ref{type0freq} (a).

For $\gamma \neq 0$, the vortex dynamics is influenced by the intercomponent interaction 
when the initial separation $R$ is small as shown in Fig.~\ref{type0freq}(a). 
Although the vortices also make a circular motion, 
its rotation frequency has a $\gamma$-dependence, as shown in Fig. \ref{type0freq}(b), 
where we plot a rotation frequency for $R=3$ as a function of $\gamma$. 
The rotation frequency is slightly increased with changing the $\gamma$ from zero to 
negative values. On the other hand, it decreases with increasing $\gamma$ to positive values. 

\subsubsection{The different circulation: $(1,0)$-$(-1,0)$}   
\begin{figure}[ht]
\centering
\includegraphics[width=0.9\linewidth,bb=0 0 540 595]{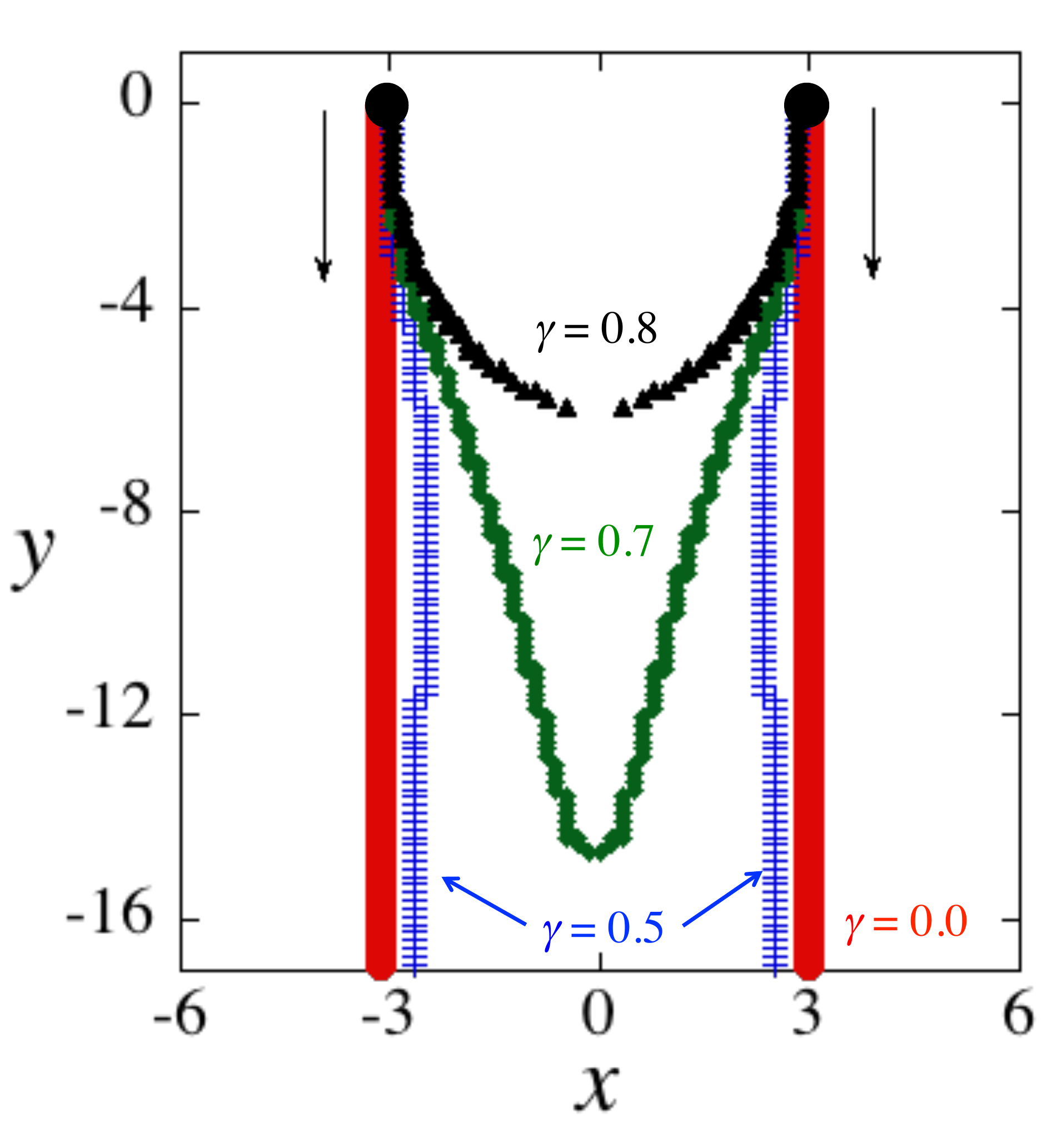} \\\
\caption{The dynamical trajectories of vortex dipoles prepared initially in the $\psi_1$-component, 
where we put a vortex at $x=R$ and an anti-vortex at $x=-R$ with $R=3$.  
The (red) circles, (blue) cross, (green) diamonds, and (black) triangles 
represent the results for $\gamma=0, 0.5, 0.7$, and $0.8$, respectively. 
}
\label{type0mtraj}
\end{figure}

Next, we prepare the initial state as $(1,0)$-$(-1,0)$, i.e., vortex dipoles in the $\psi_1$-component, and 
simulate the vortex trajectories for several $\gamma$, which are shown in Fig.~\ref{type0mtraj}. 
For $\gamma = 0$ as well as a single-component BEC, they move  to the same direction by 
keeping its separation between them.
For $\gamma < 0$, they move by a similar way as seen in $\gamma=0$ (not shown).  
However a significant difference can be 
seen for $\gamma \geq 0.7$. Although the two vortices move to the same direction, 
they approach to each other and eventually collide 
to cause pair annihilation, as seen in Fig.~\ref{type0mtraj}. 
For all the values of $\gamma$, the initial velocity of the vortices does not 
change so much, which can be seen in Fig.~\ref{type0rot}(b) 
in which the curvature 
of the potential for different $\gamma$ does not change around $R=3$.

\subsection{Analysis by point vortex approximation}\label{analytis}
When we apply the simple Abrikosov ansatz of Eq.~(\ref{initialhqv0}) 
and the adiabatic approximation for the wave function,
we can derive the Euler-Lagrange equations similar to Eqs.~(\ref{vpmotion01})-(\ref{vpmotion04}).
Here, the intervortex potential $V_{12}$ should be replaced by that shown in Fig.~\ref{type0rot} for $(1,0)$-$(\pm1,0)$ case.  
The typical solution has the same form with Eq.~(\ref{unicycmo0}) for $(1,0)$-$(1,0)$
and Eq.~(\ref{unilinmo0}) for $(1,0)$-$(-1,0)$.
Since the intervortex potential has a $\gamma$-dependance within a short range of $R$ as in Fig.~\ref{type0rot}, 
we expect some quantitative change of the velocity of motion. 
The rotation frequencies for $R=3$ obtained by the intervortex potential $V_{12}$ are shown 
in Fig.~\ref{type0rot}(b). 
The two methods, simple- and improved-Abrikosov ansatz can predict correct 
rotation frequency for the range $-1 \leq \gamma \leq 0.2$. As $\gamma$ increases further, 
the simple Abrikosov ansatz yields unreliable result for $\gamma \geq 0.7$, which 
means the breakdown of this, see ansatz Fig.~\ref{type0freq} (c). 
The improved method can reproduce the 
decreasing behavior even for $\gamma > 0.7$, but the result is not agreed 
quantitatively. Although there is a local minimum in $V_{12}$ around $R=0$ for $\gamma > 0.7$ in 
the latter method, the numerical simulations show that two vortices do not 
exhibit such an attracting behavior even at $\gamma = 0.9$ \cite{tyu2}. 
Thus, the improved ansatz Eqs.~(\ref{impabrianz1}) and (\ref{impabrianz2})
is better than the simple Abrikosov ansatz but it cannot be applicable for $0.7 < \gamma < 1$.
Since Eqs.~(\ref{impabrianz1}) and (\ref{impabrianz2}) only consider the 
deformation of the amplitude of the wave function, more proper ansatz including 
the deformation of the phase field might improve the quantitative prediction. 

\section{Summary and discussion} \label{summary}
In this paper, we have discussed the intervortex interaction and the dynamics of 
two HQVs in two-component BECs. 
The direct numerical simulations of the GP equations reveal nontrivial vortex dynamics 
as shown in Fig.~\ref{a12=-0.5}-\ref{a12=0.5m}, which are quite different 
from those in a single-component BEC. 
The intervortex interaction can be represented 
very well by the asymptotic form derived by Eto \textit{et al}. even when the vortices 
are not far from each other.
In our model, the vortex dynamics can be described by a balance between 
the Magnus force and the intervortex force. 
Through the proper treatment to estimate the intervortex force, the initial stage of the 
observed dynamics can be explained by this picture. 

However, some of the qualitative behaviors of the vortex motion have not been explained yet. 
In the $(1,0)$-$(0,\pm1)$ case, since the vortices tend to go away from or approach to each other when the 
intervortex interaction is repulsive or attractive, respectively, the dynamics is 
similar to those of vortices in relativistic theory \cite{Manton}, where the time derivative of 
the equation of motion is second order. 
If there are second-order time derivative terms in the left-hand-sides of 
Eqs.~(\ref{vpmotion01})-(\ref{vpmotion04}), whose coefficients may be 
refereed to as ``effective vortex mass", vortex dynamics can obtain 
an acceleration caused by intervortex force. We need more consideration to clarify 
whether the problem is solved by including these terms, and how these terms 
can be derived. 
Note that Nakamura {\it et al.} obtained the vortex mass by introducing the additional 
degrees of freedom in the phase of the condensate wave function \cite{Nakamura}. 
A proper treatment of the phase dynamics might give rise to the 
expected mass term in the equation of motion. 

Also for $(1,0)$-$(-1,0)$ case, the vortex point model cannot explain the 
the approaching trajectories of the vortices seen in Fig.~\ref{type0mtraj}.
We may have some conjectures from the numerical observations. 
As seen in Sec.~\ref{numety2}, it is likely that the vortices tend to approach 
due to the attractive interaction between the vortices. This fact implies that 
the vortex might get an effective mass, which involves the acceleration of 
the motion. Another possibility is that the vortex motion seems to be dissipated 
by some mechanism. In the absence of dissipation, the vortex dipole 
should move by keeping their separation and this is actually observed for 
$\gamma \leq 0$. For $\gamma>0$ and even in the energy conserved system, 
effective mutual friction may occur in counterflowing 
two-component superfluids due to the momentum exchange between two components \cite{Takeuchi}.
Then, the vortex energy can escape to a kind of compressible energy in the other component. 

Since the intervortex force in two-component BECs is very weak, the vortex dynamics 
shown in this paper could be observable for vortices with a relatively short distance.   
The number density $10^{13}$-$10^{15}$ cm$^{-3}$ in typical cold atom experiments 
yields $\xi \sim 0.1$ $\mu$m and $\tau \sim 0.1$ msec. Thus, for the vortex separation 
$R \leq 5$ simulated in this paper the typical time scale of the vortex dynamics 
such as the period of the circular motion is order of $10^{2-3}$ msec. 
Much longer time is needed to detect the dynamics of initially well-separated vortices 
with $R \geq 6$; for example, according to Fig. \ref{inivelocom}(a), the period 
of the circular motion is about 2 sec for $R=6$ and 6 sec for $R=8$. 
Also, when the condensate is trapped by a harmonic potential, the vortex exhibits 
a precession motion around the center due to the density inhomogeneity \cite{Fetterrev}. 
This frequency is proportional to the inverse of the square of the Thomas-Fermi radius. 
It is necessary to prepare condensates with large Thomas-Fermi radius to 
verify our prediction. 

Half-quantized vortices in two-component BECs 
are predicted to exhibit exotic vortex lattice structures 
in the presence of the coherent Rabi coupling between the two components
\cite{Kasarev,Cipriani:2013nya}. 
The interaction and dynamics of half-quantized vortices in the presence 
of the Rabi coupling are also an important future problem. 

Generalization of our result to fractional vortices in multi-component more than two 
components \cite{Cipriani:2013wia,Eto:2012rc}  
is straightforward.  
Finally, the method introduced in this paper should be applicable to 
different systems such as fractional non-Abelian vortices in 
dense quark matter \cite{Eto:2013hoa}, 
for which the interaction between 
two well-separated vortices is the same with 
that of superfluid vortices \cite{Nakano:2007dr}.
The short range interaction is important 
to determine the vortex lattice structure \cite{Kobayashi:2013axa}.

\acknowledgements
The work of K. K. is supported by KAKENHI from the Japan Society for the Promotion of Science (JSPS) 
Grant-in-Aid for Scientific Research (KAKENHI Grant No. 26400371). 
The work of M. E. is supported in part by JSPS Grant-in-Aid for Scientific Research 
(KAKENHI Grant No. 26800119).
The work of M. N. is supported in part by 
JSPS Grant-in-Aid for Scientific Research
(KAKENHI Grant No. 25400268) and 
by a Grant-in-Aid for Scientific Research on Innovative Areas
``Topological Materials Science"
(KAKENHI Grant No. 15H05855) and 
``Nuclear Matter in Neutron Stars Investigated by Experiments and
Astronomical Observations"
(KAKENHI Grant No. 15H00841) 
from the the Ministry of Education, Culture, Sports, Science (MEXT) of Japan.
The work of M. E. and M. N. is also supported in part  by 
the MEXT-Supported Program for the Strategic Research Foundation
at Private Universities ``Topological Science" (Grant No. S1511006).

\appendix
\section{Derivation of the equation of motion for vortex coordinates}\label{appa}
Here, we describe a derivation of the equation of motion for the 
vortex coordinates. For the $(1,0)$-$(0,\pm1)$ case in Sec.~\ref{case2}, 
the wave functions for the two-component BECs are approximately 
written as Eqs.~(\ref{anaanz1})-(\ref{anaanz5}).
The time evolution of the system is only included in the coordinates $\mathbf{r}_1(t)$ 
and $\mathbf{r}_2(t)$ in the adiabatic limit. 
The Langrangian is given by Eq.~(\ref{dimlessanalag}), and the total 
energy $E_0^{(q_1,q_2)}$ [Eq.~(\ref{dimlessanaene0})] is reduced to 
the vortex-vortex interaction $V_{12}$ shown in Fig.~\ref{type12vorpot}
after substituting Eqs.~(\ref{anaanz1})-(\ref{anaanz5}) and making the integration. 

By taking the variation of the action $S = \int dt L$ with respect to $\mathbf{r}_i$, 
we obtain   
\begin{align}
\delta S = & - \sum_{i} \int dt \biggl\{ \int d \mathbf{r} \biggl[ \dot{\theta}_i (\nabla_{i} \rho_i)  
+ \rho_i (\nabla_{i} \dot{\theta}_i)  \nonumber \\  
& + \dot{\theta}_{\bar{i}} (\nabla_{i} \rho_{\bar{i}}) \biggr]
+ \nabla_{i} V_{12} \biggr\} \cdot \delta \mathbf{r}_i  \nonumber \\  
& - \sum_{i} \int dt  \int d \mathbf{r} \rho_i (\nabla_{i} \theta_{i}) \cdot \delta \dot{\mathbf{r}}_i,
\label{totyuud}
\end{align}
where $i=1,2$, $\bar{i}=2(1)$ when $i=1(2)$, and $\nabla_i \equiv \nabla_{\mathbf{r}_i}$. 
After integrating partly, the last term in Eq.~(\ref{totyuud}) becomes 
\begin{equation}
\int dt \int d\mathbf{r} ( \dot{\rho}_i \nabla_{\mathbf{r}_i} \theta_i + \rho_i \nabla_{\mathbf{r}_i} \dot{\theta}_i ) \cdot \delta \mathbf{r}_i. \label{inte2}
\end{equation}
Thus, the second term of Eq.~(\ref{totyuud}) and that of Eq.~(\ref{inte2}) are 
canceled out, so that we get 
\begin{align}
\delta S = &- \sum_{i} \int dt \biggl\{ \int d \mathbf{r} \biggl[ \dot{\theta}_i (\nabla_{i} \rho_i) 
- \dot{\rho}_i (\nabla_{i} \theta_{i})   \nonumber \\ 
& +  \dot{\theta}_{\bar{i}} (\nabla_{i} \rho_{\bar{i}}) \biggr] 
 + \nabla_{i} V_{12} \biggr\} \cdot \delta \mathbf{r}_i.
\label{totyuud2}
\end{align}
According to Eqs.~(\ref{anaanz2})-(\ref{anaanz5}), we have the relation 
\begin{align}
\dot{\rho}_i &= (\nabla_{i} \rho_i) \cdot \dot{\mathbf{r}}_i + (\nabla_{\bar{i}} \rho_i) \cdot \dot{\mathbf{r}}_{\bar{i}},  \nonumber \\
\dot{\theta}_i &= (\nabla_{i} \theta_i) \cdot \dot{\mathbf{r}}_i, 
\end{align}
and $\delta S$ can be written as  
\begin{align}
\delta S = & - \sum_{i} \int dt \biggl\{ \int d \mathbf{r} \biggl[  (\nabla_{i} \theta_i) \cdot \dot{\mathbf{r}}_i  (\nabla_{i} \rho_i) \nonumber \\
& - (\nabla_{i} \rho_i) \cdot \dot{\mathbf{r}}_i  (\nabla_{i} \theta_{i}) 
 - (\nabla_{\bar{i}} \rho_i) \cdot \dot{\mathbf{r}}_{\bar{i}} (\nabla_{i} \theta_{i}) \nonumber \\
 & + (\nabla_{\bar{i}} \theta_{\bar{i}}) \cdot \dot{\mathbf{r}}_{\bar{i}}  (\nabla_{i} \rho_{\bar{i}})   \biggr]  
+ \nabla_{i} V_{12} \biggr\} \cdot \delta \mathbf{r}_i.
 \label{totyuud2s}
\end{align}
By using the formula $(\mathbf{A} \times \mathbf{B}) \cdot (\mathbf{C} \times \mathbf{D}) 
= (\mathbf{A} \cdot \mathbf{C})  (\mathbf{B} \cdot \mathbf{D}) - (\mathbf{A} \cdot \mathbf{D})  (\mathbf{B} \cdot \mathbf{C}) $, 
the first two terms can be written as
\begin{align}
\int d\mathbf{r} (\nabla_{i} \rho_i \times \nabla_{i} \theta_i) \cdot (\delta \mathbf{r}_i \times \dot{\mathbf{r}}_i ). \label{etosankizuku}
\end{align}
Here, we note that $\nabla_{\mathbf{r}_i}$ can be replaced by $-\nabla$ when it acts on $\mathbf{r} - \mathbf{r}_i$. 
Then, $\nabla_{i} \rho_i = - \nabla \rho_i - \nabla_{\bar{i}} \rho_{i}$ and $\nabla_i \theta_i = - \nabla \theta_i$. 
Equation (\ref{etosankizuku}) becomes
\begin{equation}
\int d \mathbf{r} \left[(\nabla \rho_i \times \nabla \theta_i) - (\nabla_{\bar{i}} \rho_i \times \nabla_{i} \theta_i) \right] 
\cdot (\delta \mathbf{r}_i \times \dot{\mathbf{r}}_i ). \label{etosankizuku2}
\end{equation}
The first term can be written as 
\begin{align}
&- \int d\mathbf{r} \left[ (\nabla \rho_i \times \nabla \theta_i) \times \dot{\mathbf{r}}_i \right] \cdot \delta \mathbf{r}_i \nonumber \\
= & - \int d\mathbf{r} \left[ \nabla \times ( \rho_i \nabla \theta_i) \times \dot{\mathbf{r}}_i \right] \cdot \delta \mathbf{r}_i .
\end{align}
The spatial integral can be calculated by the Stokes theorem as
\begin{equation}
-  2\pi q_i \rho_{r=\infty} \hat{\mathbf{z}} \times \dot{\mathbf{r}}_i \cdot \delta \mathbf{r}_i.
\end{equation}
This term represents the well-known Magnus force. 
It should be noted that we still remain some integrals: (i) the third and fourth terms of Eq.~(\ref{totyuud2s}) 
\begin{align}
 \int d\mathbf{r} \biggl\{ [(\nabla_{\bar{i}} \theta_{\bar{i}}) \cdot \dot{\mathbf{r}}_{\bar{i}} ]  (\nabla_{i} \rho_{\bar{i}} ) 
 - [ (\nabla_{\bar{i}} \rho_i) \cdot \dot{\mathbf{r}}_{\bar{i}} ] ( \nabla_{i} \theta_i )  \biggr\} \cdot \delta \mathbf{r}_i , 
 \label{addcontriun}
\end{align}
and (ii) the second term of Eq.~(\ref{etosankizuku2})
\begin{align}
 \int d\mathbf{r} \biggl\{ [(\nabla_{i} \theta_{i}) \cdot \dot{\mathbf{r}}_{i} ]  (\nabla_{\bar{i}} \rho_{i} )  
 - [ (\nabla_{\bar{i}} \rho_i) \cdot \dot{\mathbf{r}}_{i} ] ( \nabla_{i} \theta_i )  \biggr\} \cdot \delta \mathbf{r}_i . 
  \label{addcontriun2}
\end{align}
These integrals consist of cross terms of the gradient of the density or the phase 
at the different position vectors, which may be a small contribution for the integration 
due to the small overlap for well separated vortices. 
As seen in the Appendix \ref{additionalc}, these terms give 
a minor contribution to the motion of point vortices without any change of the 
qualitative behaviors. 

Since $\rho_{r=\infty} = 1$ in our unit, the equation of motion is written by 
\begin{align}
&2 \pi q_i \dot{\mathbf{r}}_i \times \hat{\mathbf{z}} = -\nabla_{i} V_{12} \nonumber \\
& + \int d \mathbf{r} \biggl[ (\nabla_{\bar{i}} \rho_i) \cdot \dot{\mathbf{r}}_{\bar{i}} (\nabla_{i} \theta_i) 
- (\nabla_{\bar{i}} \theta_{\bar{i}}) \cdot \dot{\mathbf{r}}_{\bar{i}}  (\nabla_{i} \rho_{\bar{i}})  \biggr] \nonumber \\
& + \int d \mathbf{r} \biggl[ (\nabla_{i} \theta_i) \cdot \dot{\mathbf{r}}_{i} (\nabla_{\bar{i}} \rho_i) 
- (\nabla_{\bar{i}} \rho_{i}) \cdot \dot{\mathbf{r}}_{i}  (\nabla_{i} \theta_{i})  \biggr].
\end{align}
By neglecting the last two integrals, we arrive at Eqs.~(\ref{vpmotion01})-(\ref{vpmotion04})

Following the same procedure, we can also derive the equations of motion 
for the $(1,0)$-$(\pm1,0)$ case. 
The ansatz is given by Eqs.~(\ref{initialhqv0}).
The variation of the action becomes 
\begin{align}
\delta S =\sum_{i} \int dt \biggl\{ \int d\mathbf{r} \left[ (\nabla_{i} \rho_1)  \dot{\theta}_1 - \dot{\rho}_1 (\nabla_{i} \theta_1) \right]  \nonumber \\
 - \nabla_{i} V_{12}  \biggr\} \cdot \delta \mathbf {r}_i.
\end{align}
Using the relations 
\begin{align} 
\dot{\rho}_1 = (\nabla_{1} \rho_1) \cdot \dot{\mathbf{r}}_1 + (\nabla_{2} \rho_1) \cdot \dot{\mathbf{r}}_2, \nonumber  \\
\dot{\theta}_1 = (\nabla_{1} \theta_1) \cdot \dot{\mathbf{r}}_1 + (\nabla_{2} \theta_1) \cdot \dot{\mathbf{r}}_2, 
\end{align}
we get 
\begin{align}
& \delta S = - \sum_{i} \int dt \biggl\{ \nabla_{i} V_{12} \nonumber \\
&+ \int d\mathbf{r} \left[ (\nabla_{i} \theta_1) \cdot \dot{\mathbf{r}}_i (\nabla_{i} \rho_1) 
- (\nabla_{i} \rho_1) \cdot \dot{\mathbf{r}}_i (\nabla_{i} \theta_1) \right] \nonumber \\
& + \int d\mathbf{r} \left[ (\nabla_{\bar{i}} \theta_1) \cdot \dot{\mathbf{r}}_{\bar{i}} (\nabla_{i} \rho_1)
-(\nabla_{\bar{i}} \rho_1) \cdot \dot{\mathbf{r}}_{\bar{i}} (\nabla_{i} \theta_1)  \right]  \biggr\}  
\cdot \delta \mathbf {r}_i. \label{act0101}
\end{align}
The integral in the second line gives the Magnus force term 
as well as the additional integral similar to Eq.~(\ref{addcontriun2}), where 
we use the replacement $\nabla_i \rho_1 = -\nabla \rho_1 - \nabla_{\bar{i}} \rho_1$ and 
$\nabla_i \theta_1 = - q_i \nabla \theta^{0}(\mathbf{r} - \mathbf{r}_i)$. 
The integral in the third line also provides additional contribution similar to Eq.~(\ref{addcontriun}). 
There is a difference from the $(1,0)$-$(0,\pm1)$ case in the additional integrals, 
where the gradient of the density is written by the core profile $f(\mathbf{r})$ 
instead of the hump profile $h(\mathbf{r})$. 
These are generally neglected in the analysis of a single-component superfluid 
\cite{Freilich,Middelkamp,Navarro} as well as our analysis in the text. 
The equation of motion is given by 
\begin{align}
& 2 \pi q_i \dot{\mathbf{r}}_i \times \hat{\mathbf{z}} = -\nabla_{i} V_{12} \nonumber \\
&  + \int d \mathbf{r} \biggl[ (\nabla_{\bar{i}} \rho_1) \cdot \dot{\mathbf{r}}_{\bar{i}} (\nabla_{i} \theta_{1})  
- (\nabla_{\bar{i}} \theta_1) \cdot \dot{\mathbf{r}}_{\bar{i}}  (\nabla_{i} \rho_1)  \biggr] \nonumber \\
& + \int d \mathbf{r} \biggl[ (\nabla_{i} \theta_1) \cdot \dot{\mathbf{r}}_i (\nabla_{\bar{i}} \rho_{1})  
- (\nabla_{\bar{i}} \rho_1) \cdot \dot{\mathbf{r}}_i  (\nabla_{i} \theta_1)  \biggr].
\end{align}

\section{Contribution of the additional integrals: Eqs.~(\ref{addcontriun}) and (\ref{addcontriun2})}\label{additionalc}
Here, we discuss the additional contributions Eqs.~(\ref{addcontriun}) and (\ref{addcontriun2}) 
shown in the previous derivation. 
We focus on the $(1,0)$-$(0,\pm1)$ vortex state 
for explanation, but it is essentially the same for the $(1,0)$-$(\pm 1,0)$ case. 

\begin{figure}[ht]
\centering
\includegraphics[width=1.0\linewidth,bb=0 0 510 481]{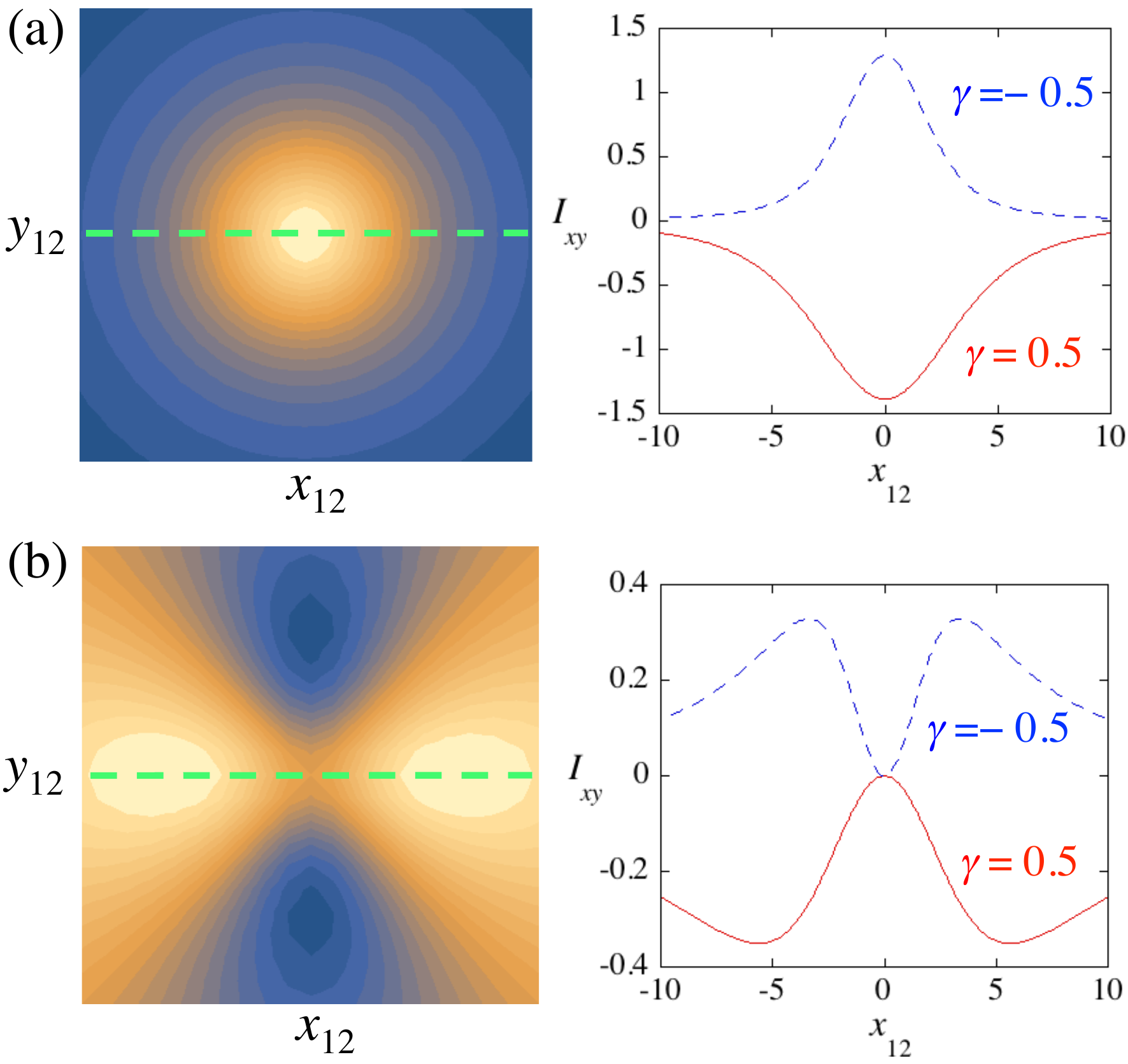} \\\
\caption{(Color online) Typical profiles of the integrals (a) $I_{xy}$ and $I_0$ ($I_{xy}=I_0$) for $(1,0)$-$(0,1)$ 
and (b) $I_{xy}$ for $(1,0)$-$(0,-1)$. The left panels show the contourplots of $I_{xy}$ for $\gamma<0$; 
for $\gamma > 0$ the sign of $I_{xy}$ is inverted from that of $\gamma<0$. 
The value is larger (smaller) in the brighter (darker) region. 
In the right panel, we plot the cross section of $I_{xy}$ along 
the $y_{12} = 0$ line for $\gamma=0.5$ and $-0.5$.}
\label{type12act}
\end{figure}

With remaining these additional terms, the equations of motion 
can be written as 
\begin{align}
(2 \pi-I_0) q_1 \dot{y}_1 &= - \frac{\partial V_{12}}{\partial x_1} - I_{xx} \dot{x}_2 - I_{xy} \dot{y}_2, \label{vpmotion1ap} \\
-(2 \pi-I_0) q_1 \dot{x}_1 &= - \frac{\partial V_{12}}{\partial y_1} - I_{yx} \dot{x}_2 - I_{yy} \dot{y}_2, \label{vpmotion2ap} \\
(2 \pi-I_0) q_2 \dot{y}_2 &= - \frac{\partial V_{12}}{\partial x_2} + I_{xx} \dot{x}_1 + I_{yx} \dot{y}_1, \label{vpmotion3ap} \\
-(2 \pi-I_0) q_2 \dot{x}_2 &= - \frac{\partial V_{12}}{\partial y_2} + I_{xy} \dot{x}_1 + I_{yy} \dot{y}_1.\label{vpmotion4ap}
\end{align}
The coefficients $I_{0}$ and $I_{\alpha \beta}$ $(\alpha,\beta=x,y)$ are given by 
\begin{align}
I_0 = \int  d\mathbf{r} \biggl[ - f(\mathbf{r}-\mathbf{r}_1) v^0_{x}(\mathbf{r}-\mathbf{r}_1) \partial_{y} h(\mathbf{r}-\mathbf{r}_2)  \nonumber  \\ 
+ f(\mathbf{r}-\mathbf{r}_1) v^0_{y}(\mathbf{r}-\mathbf{r}_1) \partial_{x} h(\mathbf{r}-\mathbf{r}_2) \biggr], \label{addactiontype00} \\
I_{\alpha \beta} = \int d\mathbf{r} \biggl[ - q_1 f(\mathbf{r}-\mathbf{r}_1) v^0_{\alpha}(\mathbf{r}-\mathbf{r}_1) \partial_{\beta} h(\mathbf{r}-\mathbf{r}_2)  \nonumber  \\ 
+ q_2 f(\mathbf{r}-\mathbf{r}_2) v^0_{\beta}(\mathbf{r}-\mathbf{r}_2) \partial_{\alpha} h(\mathbf{r}-\mathbf{r}_1) \biggr] \label{addactiontype12} 
\end{align}
with $v_{\alpha}^0 (\mathbf{r} - \mathbf{r}_i) = \partial_{\alpha} \theta^{0}(\mathbf{r}-\mathbf{r}_i)$. 
The integrals $I_0$ and $I_{\alpha \beta}$ come from Eqs.~(\ref{addcontriun2}) and (\ref{addcontriun}), respectively. 
These are invariant under the change $q_1 \leftrightarrow q_2$ and 
$\mathbf{r}_1 \leftrightarrow \mathbf{r}_2$ 
and functions of the relative position $\mathbf{r}_{12} = \mathbf{r}_2-\mathbf{r}_1 =(x_{12},y_{12})$, 
which can be easily seen by changing the variables as $\mathbf{r} \to \mathbf{r} - \mathbf{r}_{12}/2$. 
The integrand consists of an overlap of the current density around $\mathbf{r}_1$ 
and the density gradient around $\mathbf{r}_2$, and that of vice versa. 
For $\gamma = 0$, these integrals vanish because of $\nabla h = 0$.
For the well separated vortices, the overlap should be small. 
For short-distance dynamics, however, these integrals may have some 
contribution. 
The typical values of $I_0$ and $I_{\alpha \beta}$ as functions of $x_{12}$ and $y_{12}$ 
for $\gamma = \pm 0.5$ are shown in Fig.~\ref{type12act}, 
which are calculated by the numerical integral with the approximated form 
Eqs.~(\ref{eq:gp_varia1}), (\ref{eq:gp_varia2}), and (\ref{phase0prof}) of the wave function. 
As seen in Eqs.~(\ref{addactiontype00}) and (\ref{addactiontype12}), $I_{xy} = I_0$ if $q_1=q_2$ 
(the second integrals of Eqs.~(\ref{addactiontype00}) and (\ref{addactiontype12}) are equivalent 
under $\mathbf{r}_1 \leftrightarrow \mathbf{r}_2$).
We remain these terms here, finding that 
there are only small correction of the quantitative dynamical 
feature without the change of the qualitative dynamics. 

Let us see the impact of these terms to the equations of motion. 
For $(1,0)$-$(0,1)$ [$(q_1,q_2)=(1,1)$], the integral becomes $I_0 = I_{xy}$, $I_{xx} = I_{yy} = 0$, 
and $I_{yx} = -I_{xy}$ because 
of the inversion symmetry $\mathbf{r}_1 \leftrightarrow \mathbf{r}_2$. Moreover, $I_{xy}$ 
is a isotropic function of $\mathbf{r}_{12}$, being written as $I_{xy} (r_{12})$. 
Since $\mathbf{r}_{12}$ is still a constant of motion even in the presence of $I_{xy}$, 
$I_{xy}$ and $I_0$ are constants determined by the initial position of the vortices. 
We obtain the equation of motion for $\mathbf{r}_i$ as
\begin{equation}
\ddot{\mathbf{r}}_i = - \left[ \frac{2}{(2\pi-I_0-I_{xy}) r_{12}} \frac{\partial V_{12}}{\partial r_{12}} \right]^2 \mathbf{r}_i. \label{unicycmo}
\end{equation}
This gives an uniform circular motion and the rotation frequency depends on 
the $\gamma$. The factor $I_0+I_{xy}$ only modifies the rotation frequency as a function of $r_{12}$.  

For $(1,0)$-$(0,-1)$ [$(q_1,q_2)=(1,-1)$], the inversion symmetry $\mathbf{r}_1 \leftrightarrow \mathbf{r}_2$ 
leads to $I_{xx} = - I_{yy}$ and $I_{yx} = I_{xy}$, but $I_0$ is not equal to $I_{xy}$. 
Even with this additional terms, one can show that $\mathbf{r}_{12}$ is a constant of motion. 
An appropriate choice of the initial condition, such as $x_{12}(0) =  2R$ and $y_{12}(0) = 0$ 
as in our simulations, makes $I_{xx}$ vanish and $I_{xy}$ being constant determined by the 
initial condition. 
The solution is given by $x_{1(2)} = (-) R$ and 
\begin{equation}
y_1=y_2=\frac{1}{2\pi-I_0-I_{xy}} \left( \frac{\partial V_{12}}{\partial r_{12}} \right) t. \label{unilinmo}
\end{equation}
This is exactly an uniform linear motion for the both vortices.

\begin{figure}[ht]
\centering
\includegraphics[width=0.9\linewidth,bb=0 0 453 567]{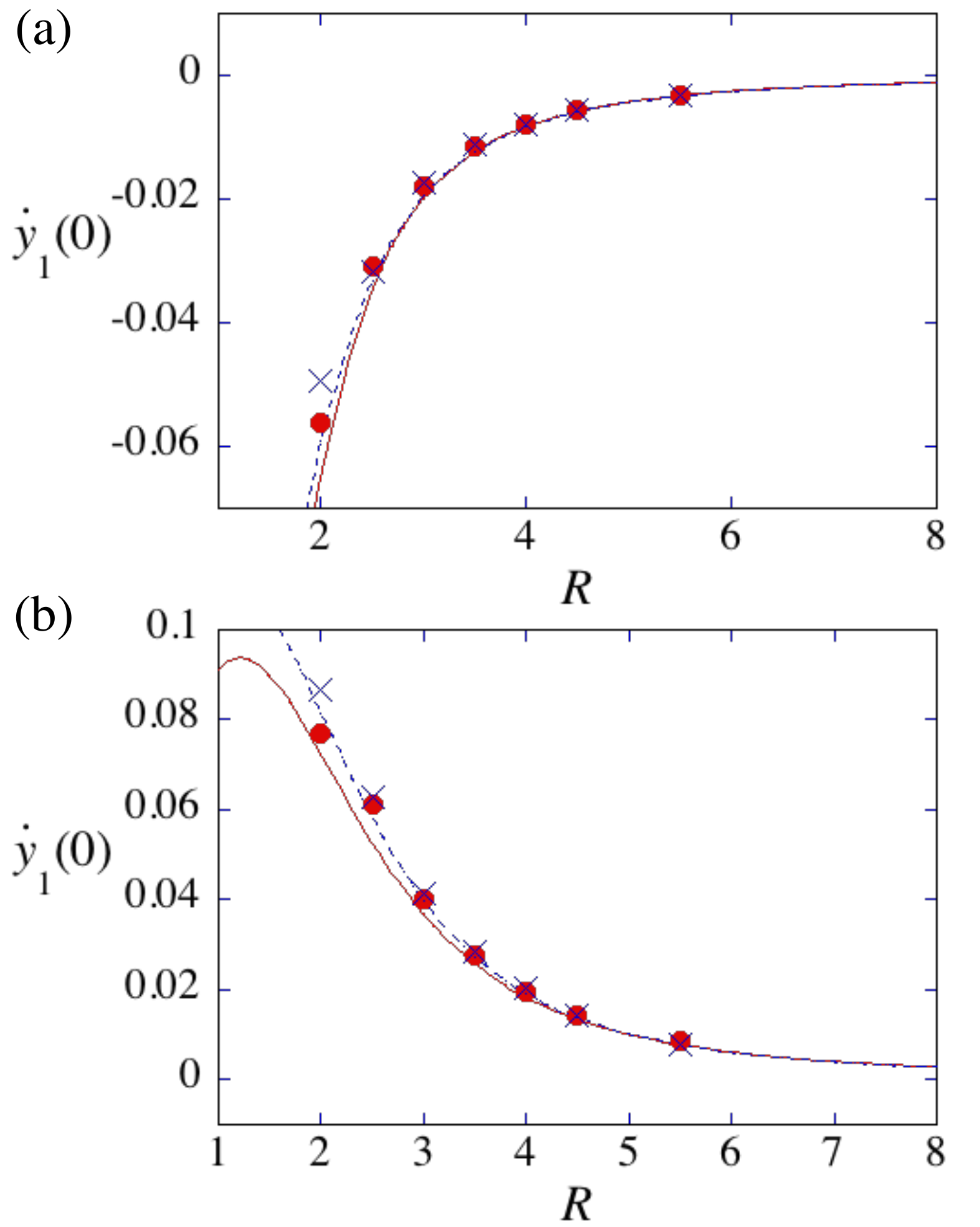} \\\
\caption{A similar plots with Fig.~\ref{inivelocom}, but the numerical results are 
compared with the initial velocity calculated from 
the intervortex force with the correction of $I_0$ and $I_{xy}$ 
 for (a) $\gamma = -0.5$ and (b) $\gamma = 0.5$.  
The (red) circles and (blue) crosses represent 
the numerical results of the GP equation for $(1,0)(0,1)$ and $(1,0)(0,-1)$, respectively.
The solid and dashed curves represent $(\partial V_{12}/\partial r_{12}) /(2 \pi-I_0-I_{xy})$ for 
$(1,0)(0,1)$ and $(1,0)(0,-1)$, respectively. 
}
\label{inivelocorr}
\end{figure}

In both cases, Eqs.~(\ref{unicycmo}) and (\ref{unilinmo}) imply that $I_{0}$ and $I_{xy}$ do not 
change any qualitative behavior of the vortex dynamics, but 
the initial velocity of the vortices can be modified through the factor $2\pi -I_0- I_{xy}$.
The quantitative correction due to $I_0$ and $I_{xy}$ is very small, except for 
$|x_{12}| < 5$ $(R<2.5)$ in the $(1,0)$-$(0,1)$ case 
where $I_{xy} \sim {\cal O}(1)$ [Fig.~\ref{type12act}(a)]. However, when we take account of 
the small correction, the initial velocity obtained from the GP simulations in Fig.~\ref{inivelocom} is 
explained better. Figure~\ref{inivelocorr} shows the initial velocity of the GP equations 
in Fig.~\ref{inivelocom} again as well as that obtained from the intervortex 
potential Eq.~(\ref{etopotori}) including the correction $I_{xy}$. The correction is included 
in a different way for $(1,0)$-$(0,1)$ and $(1,0)$-$(0,-1)$ case as seen in Fig.~\ref{type12act}; 
for $\gamma>0$ the initial velocity of $(1,0)$-$(0,1)$ is slightly smaller than that of $(1,0)$-$(0,-1)$, 
and vice versa for $\gamma<0$. This is actually observed in the numerical results. 

For the $(1,0)$ and $(\pm1,0)$ case, the integrals $I_{0}$ and $I_{\alpha \beta}$ are given by 
\begin{align}
I_{0} = \int d\mathbf{r} \biggl[ -  f(\mathbf{r}-\mathbf{r}_1) v^0_{x}(\mathbf{r}-\mathbf{r}_1) \partial_{y} f(\mathbf{r}-\mathbf{r}_2)  \nonumber  \\ 
+ f(\mathbf{r}-\mathbf{r}_1) v^0_{y}(\mathbf{r}-\mathbf{r}_1) \partial_{x} f(\mathbf{r}-\mathbf{r}_2) \biggr],  \nonumber \\
I_{\alpha \beta} = \int d\mathbf{r} \biggl[ - q_1 f(\mathbf{r}-\mathbf{r}_1) v^0_{\alpha}(\mathbf{r}-\mathbf{r}_1) \partial_{\beta} f(\mathbf{r}-\mathbf{r}_2)  \nonumber  \\ 
+ q_2 f(\mathbf{r}-\mathbf{r}_2) v^0_{\beta}(\mathbf{r}-\mathbf{r}_2) \partial_{\alpha} f(\mathbf{r}-\mathbf{r}_1) \biggr].  \label{addactiontype0} 
\end{align}
The difference from Eq.~(\ref{addactiontype12}) is that the density gradient 
is altered by the vortex core profile $f(\mathbf{r})$ instead of the hump 
profile $h(\mathbf{r})$. 
These coefficients can be also 
calculated by the numerical integrals, and the overall profile is consistent with 
those in Fig.~\ref{type12act}, except its sign. The above argument is also applicable 
to this situation because the equation of motion is essentially identical to 
Eqs.~(\ref{vpmotion1ap})-(\ref{vpmotion4ap}).


\begin{thebibliography}{99}
\bibitem{tyuu}
We made similar calculations by using the Dirichlet boundary condition 
and confirmed that the results are same as those done by the Neumann condition.
We also confirm that the boundary effect is negligible by seeing 
that a single vortex at some position does not move.
\bibitem{tyu2}
We confirmed that for $\gamma = 0.9$ two initially close vortices move so as to repel each other 
through the imaginary time evolution.
\end{thebibliography}
\end{document}